\documentclass[preprint,12pt,showpacs,aps,amsmath,amssymb, 
nofootinbib,superscriptaddress,showkeys]{revtex4} 

\usepackage[toc,page]{appendix}

\usepackage{epsfig} 
\usepackage{wasysym} 
\usepackage{mathrsfs} 
\usepackage{graphicx} 
\usepackage{amsfonts} 
\usepackage{amsbsy} 
\usepackage{amscd} 
\usepackage{pstricks} 
\usepackage{enumerate}
\usepackage{color}
\definecolor{myred}{rgb}{0.6,0,0} 
\definecolor{myblue}{rgb}{0,0.2,0.4}
\definecolor{mygreen}{rgb}{0,0.9,0.1}
\definecolor{hc}{rgb}{.9,0.1,0.7}
\definecolor{hcout}{rgb}{.9,0.7,0.9}
\definecolor{Orange}{rgb}{1.,0.65,0.}

\usepackage{array}
\newcolumntype{C}[1]{>{\centering\arraybackslash$}p{#1}<{$}}

\usepackage{bm} 
\newcommand{\be}{\begin{equation}}
\newcommand{\ee}{\end{equation}}
\newcommand{\bes}{\begin{equation*}}
\newcommand{\ees}{\end{equation*}}
\newcommand{\bea}{\begin{eqnarray}}
\newcommand{\eea}{\end{eqnarray}}
\newcommand{\beas}{\begin{eqnarray*}}
\newcommand{\eeas}{\end{eqnarray*}}
\newcommand{\Tr}{\text{Tr}}
\newcommand{\la}{\lambda}

\newcommand{\Band}{\bm{\And}}
\newcommand{\Bor}{\bm{||}}
\newcommand*{\bord}{\multicolumn{1}{c|}{}}


\begin{document}

\author{Joydeep Chakrabortty}
\email{joydeep@iitk.ac.in}
\affiliation{Department of Physics, Indian Institute of Technology, Kanpur-208016, India} 

\author{Partha Konar}
\email{konar@prl.res.in}
\affiliation{Physical Research Laboratory, Ahmedabad-380009, India}

\author{Tanmoy Mondal}
\email{tanmoym@prl.res.in}
\affiliation{Physical Research Laboratory, Ahmedabad-380009, India}

\title{Copositive Criteria and Boundedness of the Scalar Potential}

\begin{abstract}
 To understand physics beyond the Standard Model (SM) it is important to have the precise knowledge of Higgs boson and 
top quark masses as well as strong coupling. Recently discovered new boson which is likely to be the SM Higgs with mass 123-127 GeV 
has a submissive impact on the stability of the new physics beyond standard model (BSM). The beyond standard model scenarios that include many scalar fields
 posses scalar potential with many quartic couplings. Due to the complicated structures of such scalar potentials it is indeed difficult
 to adjudge the stability of the vacuum. Thus one needs to formulate a proper prescription for computing the vacuum stability criteria. In 
this paper we have used the idea of copositive matrices to deduce the conditions that guarantee the boundedness of the scalar potential. 
We have discussed the basic idea behind the copositivity and then used that to determine the 
vacuum stability criteria for the Left-Right symmetric models with doublet, and triplet scalars and Type-II seesaw. As this idea is based on 
the strong mathematical arguments it helps to compute simple and unique stability criteria embracing the maximum allowed parameter space.
\end{abstract}



\maketitle

\section{Introduction} \label{sec:intro}
 
 ATLAS \cite{Aad:2012tfa} and CMS \cite{Chatrchyan:2012ufa} both have announced 
the existence of a new boson within the mass range 123-127 GeV. The trend of data hints that this might finally be the 
long sought Standard Model (SM) Higgs boson and undoubtedly it would further propel the search for new physics beyond the Standard Model (BSM).

The Higgs quartic coupling, $\lambda_h$, can be deduced solely from the Higgs mass within the SM. 
It is well known that the electroweak scalar potential is bounded from below iff the quartic coupling is non-negative, i.e., $\la_h \geq 0$. 
Thus $\la_h$ must be non-negative all the way till the Planck scale modulo the fact that there is no other theory than the SM.
The renormalisation group evolution 
of $\lambda_h$ limits two boundary values at the electroweak scale corresponding to  
the values $\pi$, and $0$ at the Planck scale. These are outcome from the demands of perturbativity of the coupling (triviality) and stability of the vacuum respectively. 
It has been noted in \cite{Degrassi:2012ry_vs,Alekhin:2012py_vs,Masina:2012tz_vs}
that the SM quartic coupling does not remain non-negative till the Planck scale for most of the parameters 
-- like, top-quark and Higgs masses, as well as, the strong coupling constant $\alpha_s$.

In spite of some aesthetic issues there are experimental evidences suggesting that SM cannot be a complete theory of nature. Thus one 
needs to look for BSM physics. As these extended models possibly with enriched group structures 
may contain exotic scalars, the potential terms are modified. Similar to the SM, new extended scalar potential must also be bounded from 
below, i.e., one needs to find the criteria for the vacuum stability\footnote{In our analysis we have used the phrases vacuum stability 
and boundedness in same footing. We have kept aside the related issues regarding the metastability for future study.}.
In this study we have revisited the old but significant question: {\it how can one ensure the vacuum stability of a scalar potential?}
It has been noted that while adjudging the vacuum stability of the scalar potential the quartic coupling parts play the most crucial role. 
In the extended scalar field models which are certainly beyond the SM, contain complicated quartic couplings. Thus it is important to 
formulate the criteria that define the boundedness of the potential 
from below and also allows the largest parameter space.  In our earlier work \cite{JD-PK-TM} we have computed some set of 
vacuum stability conditions following a technique discussed in \cite{Arhrib:2011uy_vs}. The underlying basic idea was to construct the 
quartic couplings as a pure square of the combinations of bilinear scalar fields and set their coefficients to be non-negative. 
That certainly makes the vacuum stable but for complicated potential structures certain amount of ambiguities arise. 
Here we have imported the idea of copositivity of symmetric matrices and 
interestingly the idea of the copositivity mimics the demand of boundedness of the scalar potential.

The idea of copositivity was mostly used in the area of optimisation and non-linear programming. However in the 
context of boundedness of scalar potential this was first discussed in \cite{Kannike_Kristjan}. 
So far the usability of this approach was restricted unto certain order ($\sim 3$) of matrices. In \cite{Kannike_Kristjan}, the models considered to 
demonstrate the usefulness of this technique are suited for lower order (upto order three) matrices 
and thus readily implementable through simple substitution of copositivity criteria. 
But in this present paper we have strengthened  this technique to deal with more complex scalar potentials revealing the
usefulness of copositivity widely. We have displayed these features considering few well-known class of models. 
Here the scalar sectors are much more richer and scalar potential contains many quartic couplings. Thus the matrices that need to be copositive to ensure the potential to be bounded from below is not restricted up to only order three.
Moreover, it has been noted earlier \cite{Kannike_Kristjan} that there appear some auxiliary parameters which are not physical, i.e., absent in the Lagrangian, while constructing the matrices of order 3 or higher. In this work, we also demonstrate a general mechanism to handle 
those unphysical parameters and also show how to get rid of them using some explicit examples. We believe the present approach can be applied to any models irrespective of the scalar structures.

We organise our paper as follows. First we discuss the definitions and basic ideas  of the copositivity ({\sc cop}) of the symmetric 
matrices which are constructed out of some quadratic forms. 
Discussing some of the important properties of these conditions, we have presented the general analytical forms such that one can
readily use them to derive the results. These {\sc cop} conditions are derived and analysed 
extensively by the mathematics community for some time.
Then we show how one can connect the copositivity criteria with the boundedness of the  scalar potential 
and guarantees the stability of the vacuum. We have explicitly computed the copositive criteria, i.e., 
the vacuum stability conditions for Left-Right symmetric theory with doublet scalars.
In the next section we have demonstrated the usefulness copositivity formalism as a case study. 
Using this formalism one can write in the stability conditions in much compact and simple forms which allow the largest parameter space.
Here we have picked up Type-II seesaw model as our example and performed a comparison with existing literature. 
Thereafter in the next section, we consider an example with very enriched scalar sectors, such as, Left-Right symmetric model with triplet scalars.
We have argued and shown that analytical forms of the {\sc cop} criteria is best described up to certain finite order of matrices. Since, the 
formalism of copositivity is still profound and applicable numerically, one can explore an alternative algorithm to check the copositivity  
of matrices of any finite order $n$ using the principal sub-matrices.  
This method does not give us the {\sc cop} criteria in analytical forms unlike the previous one rather check the copositivity numerically.  
In the appendix we first briefly discuss the explicit copositivity criteria in analytic forms for matrices of orders two, three and four.
Then we have computed the analytic forms of the {\sc cop} criteria for above mentioned three models. Then at the end, we have provided a brief 
algorithm with a numerical example to test copositivity of any symmetric matrix of finite order.

\section{Copositivity of Symmetric Matrix}

The concept of copositivity was first proposed as early as 1952 \cite{Motzkin} and thereafter many literatures 
in mathematics discussed and generalised the idea, see e.g. some of the widely accepted references \cite
{Hadeler198379, Väliaho198619}. It has been noted that the positive definite matrices are subset of the 
copositive (conditionally positive) matrices and are included within that. 
Some of the references, extremely  useful for our analysis, in which copositive criteria has been discussed 
profoundly are \cite{Li_Feng, kaplan_1, Kaplan_2} and more recently in \cite{bundfuss2009copositive}. 

 To define the copositivity, let us consider a set of symmetric matrices $\mathcal{S}_n$ of order $n$, and a real coordinate vector space $\Re_{n}$. 
{\it Any matrix $\Lambda \in \mathcal{S}_n$ is defined to be copositive iff the quadratic form $x^T \Lambda x \geq 0$ for 
all non-negative vectors} ($x$), {\it i.e., $x \in \Re^{+}_{n}$, the  non-negative 
orthant\footnote{An orthant can be understood as the $n$-dimensional Euclidean space, generalisation of a quadrant or an octant 
in two or three dimensions respectively.} of $\Re_{n}$}. 
It can be readily recognised the fruitfulness of explicit set of analytic conditions for a given matrix of order 
$n$ which are necessary and sufficient 
to claim the quadratic form $x^T \Lambda x \geq 0$ is copositive. 
For matrices of order two copositive conditions are straightforward and can be understood even by writing squared 
form of this quadratic expression. We have encrypted them in appendix~\ref{app:cop-order-2}. 
Although these criteria are extremely complex and long for larger order matrices with negative off-diagonal elements\footnote{Here we should keep in mind that if some of the off-diagonal 
elements of a higher order ($\geq 4$) matrix are non-negative then the copositivity can be addressed using the knowledge of {\sc cop} criteria for rather lower order matrices.}, the general {\sc cop} conditions up to order four is available, see \cite{Li_Feng}.
For the benefit of readers we also described all criteria for orders three and four in appendices~\ref{app:cop-order-3} and \ref{app:cop-order-4}.
To derive the analytic criteria for copositivity of an higher order matrix, one can make use of the rank reduction 
theorem as discussed in \cite{Li_Feng} and use the reduced rank criteria successively.
Using this rank reduction theorem, copositivity criteria for order five matrices has also been computed in \cite{order_5}.
In our analysis higher order (more than four) matrices contain many non-negative off-diagonal elements. Thus the knowledge of copositivity of order four matrix is sufficient
to deal with all of our example models.

Since analytic form of copositive criteria for the most generic symmetric matrices of larger order is not straight forward 
and not available till date, one can suitably explore the alternative algorithm exploiting the copositivity conditions recursively.
There are few approaches available in this direction and they are in principle available to incorporate numerically. We have discussed one such proposition\footnote{Yet another prescription given in \cite{Cottle1970295} puts conditions on determinant and adjugate matrix 
to find whether a matrix of order $n$ is copositive or not.} and show how they can be incorporated and tested in a numerical code.  For that purpose,
we note the theorem in \cite{kaplan_1, Kaplan_2} where it was shown that the necessary and sufficient conditions for 
the copositivity can be expressed such that, {\it $\Lambda$ is copositive iff every principal sub-matrix $\Lambda^{'}$ of $\Lambda$
does not posses any positive eigenvector associated with a negative eigenvalue}. We have extensively used this theorem to describe the algorithm as shown in appendix~\ref{app:Algo-PS} which verifies whether a given matrix of finite order is copositive or not.

Here we would like to make an important note related to the {\it basis dependency} of the copositivity of a
symmetric matrix. In general it is indeed possible to switch on some norm preserving rotation and that might change the 
structure of the quadratic form in new basis. This choice can even be made such that the matrix out of the transformed quadratic form is no longer 
copositive. But this is no eyebrow raising since any arbitrary norm preserving rotation can import the elements of 
$\Re^{-}_{n}$ and the transformed matrix might be non-copositive. Moreover, this is not the reason of worry in our case. As soon 
as we mention that this quadratic form is a part of our Lagrangian, arbitrary rotations are prohibited. From the symmetry 
principle we can allow only those rotations which leave the Lagrangian invariant. This simply says that the all allowed 
rotations will leave the quadratic form intact in any new basis. From the point of copositivity the quadratic form is the 
fundamental one, i.e., if the matrix corresponding to a quadratic form is copositive then any other matrices written in any 
other basis out of that quadratic form will be copositive.

Another important property of copositivity  is its {\sl invariance under the operation of permutation and scaling}. 
So, if $\Lambda$ be a copositive matrix then after combined operation, the new matrix $P D \Lambda D P^T$ is also 
copositive given that $P$ be  a permutation matrix and $D$ is a diagonal matrix with non-negative elements. 
We will use this property in our analysis to ensure a single set of copositivity conditions independent to the order 
of the basis elements.




\section{Copositivity and Vacuum Stability of Scalar Potential}

In the previous section we have sketched the mathematical foundation and the criteria of copositivity  
of symmetric matrices of finite order. Here we intend to implement this idea to adjudge the stability of the scalar potential.  
We will see how the copositivity guarantees the boundedness ($\geq 0$) of the quadratic form. Here we have translated the criteria 
of vacuum stability to the boundedness of the scalar potential such that we can implement the idea of copositivity. 
Since part of the scalar potential that contains the quartic couplings plays the crucial role while deciding the vacuum stability of 
corresponding potential, we shall concentrate on those terms only. We would like to note that this part of the scalar potential is 
treated as the quadratic form while the basis are bi-linear in fields. 
We construct symmetric matrices in terms of monomial basis so that the quadratic form $x^T \Lambda x \geq 0$ as expressed in the 
copositivity definition can be achieved. Hence one can directly apply the copositivity 
conditions for these $\Lambda$ matrices whose elements are made of quartic couplings. 
Thus by implementing the mathematical idea of copositivity we can guarantee the boundedness of the potential along all the field directions.  
This construction also allows us to find the largest parameter space over which the vacuum is stable. We have picked up the 
Left-Right symmetric model with doublet scalars as our  first example where we have shown how the idea of copositivity can be 
extensively used without any ambiguity. In the following section we reconsider the Type-II seesaw model and compute the copositivity 
conditions that ensure the boundedness of the potential. We also demonstrate the clarity and usefulness of this method over 
the procedure that relies on the successive squaring in different field directions as shown in \cite{Arhrib:2011uy_vs}. Then we consider a more complicated potential structure with several scalar components as suggested within  Left-Right (LR)
symmetric theory with triplet scalars. In the appendix, we have calculated the copositivity, i.e., the vacuum stability criteria for those models. There we have listed all the matrix forms of scalar potentials for different set of field directions and 
the respective criteria for copositivity  are discussed in detail. We have also shown that the matrix forms can be degenerate for different field directions. Thus the COP criteria for those cases remain same which clarifies how the basis independency of finding copositivity can be realised 
and remove unnecessary parts from the calculations. Readers who are familiar with this technique may find that the appendices \ref{app:vac_LR_doublet_COP}, \ref{app:vac_Type_II}, \ref{app:vac_LR_triplet_COP} are long. We advise them to skip these full lists and to consider 
only those field directions we referred during our discussions. Nevertheless for the benefit of curious readers we keep the full list.


\subsection{LR Model with Doublet Scalars}\label{sec:vac_LR_doublet}\label{sec:vac_LR_doublet}
 The model under consideration is as suggested in \cite{Holthausen:2011aa_smRGE} having an extended gauge symmetry which reads as $SU(3)_C\otimes SU(2)_L \otimes SU(2)_R \otimes U(1)_{B-L}$. Here the neutrino mass can be generated by a double seesaw mechanism.
Apart from a bi-doublet scalar ($\Phi$), this model contains two scalar doublets ($H_{L/R}$). The scalar potential is given in appendix \ref{app:vac_LR_doublet}. The field contents of bi-doublet scalar ($\Phi$) and $H_{L/R}$ is given as,\\
\be
\Phi =\left( \begin{array}{lr}
         \phi_1^0 \;&\; \phi_1^+\\ \\
         \phi_2^- & \phi_2^0
        \end{array} 
        \right)  \hskip 10pt , \hskip 10pt      
H_{L/R} = \left( \begin{array}{c}
                        h_{L/R}^0 \\ \\
                         h_{L/R}^+
                       \end{array}\right) .
\ee
After the neutral components of $\Phi$ and $H_{L/R}$ acquire the vacuum expectation values, the $vev$s can be represented as :
\be
\left< \Phi \right>  =\left( \begin{array}{cc}
                        v_1 & 0\\
                        0      & v_2 
                       \end{array}\right) , \hskip 20 pt                       
\left< H_L \right>  =\left( \begin{array}{c}
                        0  \\
                        v_L
                       \end{array}\right) , \hskip 20 pt                       
\left< H_R \right>  = \left(\begin{array}{c}
                        0   \\
                        v_R 
                       \end{array} \right).
\ee
Here we set $v_2=v_L=0$ to simplify the situations but without loss of any generality. The other $vev$s $v_R$ and $v_1$ represent the $SU(2)_R\otimes U(1)_{B-L}$ and $SU(2)_L \otimes U(1)_Y$ breaking scales respectively.

In appendix~\ref{app:vac_LR_doublet_COP} we have calculated the conditions for copositivity or in other words conditions for vacuum 
stability for Left-Right symmetric model with doublet scalars. 
Here we have first expanded the full scalar potential in terms of the component fields\footnote{We have considered only those scalar
multiplets whose components acquire the non-zero vacuum expectation values. 
Our assumption simplifies the analysis but does not spoil the spirit of our formalism. The most general structure surely envelopes all 
field directions but that might weaken the clarity of this formalism.}. Then we construct the quadratic forms considering 2-, 3-, and 
4-fields directions. As there are maximum four component fields ($\phi_1^0,\phi_1^+,h_R^0,h_R^+$) in our analysis all the possible 
quadratic forms are exhausted. So the copositivity has been used to its supreme without leaving any room for ambiguity. 

One can easily follow the copositivity conditions derived from the given quadratic form of the potential written in a 
symmetric matrix. These conditions directly follow from our previous discussion on general matrix. At this point, 
we would like to note and discuss some interesting situations which arise during these calculations. 
Notice that while constructing the matrix form in some of cases, we have to introduce one or more extra unphysical 
parameters to accommodate the most general multi-field terms of the potentials. 
For instance, if we consider the quadratic form like \ref{eq:vac_LR_doublet_4field} we have to construct a matrix  
that contains such new parameters\footnote{ These situations arise when some of the fields appear in 
linear form in the scalar quartic terms. They also create another problem due to the 
linearity in one of the fields. All the basis elements are not guaranteed now to belong in $\mathcal{R}_n^+$, 
which can be taken care of by introducing suitable phases in the respective couplings. 
 Now in general stability of the scalar potential can be ensured by demanding 
copositivity of the matrix form of the potential. However, one may not get a simpler set of conditions 
to ensure stability of the scalr potential as we have obtained in our model. This is the generic problem while dealing 
 with multi component scalar field models,  see ref.~\cite{Kannike_Kristjan} for example where author had discussed 
most general Two Higgs Doublet Model(2HDM). An alternative approach for this 2HDM can be found in ref.~\cite{Maniatis:2006fs,Ivanov:2006yq}.
}
in the form of $C$ and $K$. That bears interesting consequences of generating nontrivial conditions for different regions of these parameters.
However, one needs to note that $C,K$ are not the physical parameter as they do not exist at the Lagrangian level. 
Thus, we expect final conditions on stability of the potentials should be independent of these parameters.
To discuss further, we rewrite the 4-field direction conditions as computed in \ref{eq:vac_LR_doublet_4field}:
$$\la_1  \geq  0 \;\;\Band\;\;  C (f_1+2\beta_1)  \geq  0 \;\;\Band\;\;  K (f_1+2\beta_1)   \geq  0 \;\;\Band\;\;  C\;K\,\left(\frac{f_1+2\beta_1}{2}\right)^2- f_1^2  \geq  0.$$  
These conditions can be examined along with additional conditions we already derived from 
2- and 3-fields directions. These existing conditions being independent of these unphysical parameters, 
put constraint over them. So, using such conditions: $\la_1 \geq 0$ and $\beta_1 \geq |f_1|/2$,
one readily notes that both $C$ and $K$ has to be non-negative. 
Hence following this argument, last condition can be rewritten as $(2\beta_1+f_1)\geq 2|f_1|/(CK)$. 
Our intention is to find the largest parameter space which is compatible with the vacuum stability. 
In other word, one can simply evade these superficial parameters involving $C$ and $K$ by fixing their 
values leading to the largest allowed parameter space.
Following this principle, the product $C K$  which can have any non-negative value is favoured
when approaches to $\infty$. Thus the largest parameter space is allowed when the constraint is given as $(2\beta_1+f_1)\geq 0$.  
Thus the final vacuum stability conditions read as $\la_1 \geq 0$ and $\beta_1 \geq |f_1|/2$.


 
\subsection{Type--II seesaw}\label{sec:vac_Type_II}

We revisit the model given in \cite{Arhrib:2011uy_vs}, and adopted the same scalar potential for having a straightforward comparison of vacuum stability conditions.
Scalar sector of this models consists of a doublet scalar ($\Phi$) and a triplet scalar ($\Delta$) with weak hypercharge +1 and +2 respectively. 
Structures of these scalars can be  written in the following form:
\begin{center}
 $
\hskip 25pt
\Phi =\left( \begin{array}{c}
         \phi^+  \\
         \phi^0 
        \end{array} 
        \right)  \hskip 10pt , \hskip 10pt       
\Delta  = \left( \begin{array}{cc}
         \delta^+/\sqrt{2} \;&\; \delta^{++} \\   
         \delta^0 \;&\; -\delta^+/\sqrt{2}
        \end{array} 
        \right) .
$
\end{center}

For the sake of completeness we have also encoded the scalar potential in appendix~\ref{app:vac_Type_II_seesaw}.
The neutral components of the  scalars acquire $vev$s  as follows: 
\begin{center}
 $
\hskip 25pt
\langle \Phi \rangle =\left( \begin{array}{c}
         0  \\
         v 
        \end{array} 
        \right)  \hskip 10pt , \hskip 10pt       
\langle\Delta\rangle  = \left( \begin{array}{cc}
         0 \;&\; 0 \\   
         v_\Delta \;&\;0
        \end{array} 
        \right) .
$
\end{center}
leading to spontaneous breaking of the symmetry.
To establish the multi-field conditions, Arhrib $et. al.$ considered the potential along any two field directions of following form
\be
V_0 = A |\phi_A|^4 +  B |\phi_A|^2 |\phi_B|^2 +  C |\phi_B|^4,
\ee
and wrote down the necessary vacuum stability conditions as
\be\label{squre}
A \ge 0, \;\; C \ge 0\;\; \textrm{and} \;\; B + 2\sqrt{A\,C} \ge 0.
\ee
Note that these conditions are simply consistent with copositivity of order two matrix as listed in appendix~\ref{eq:copositive-matrix-order-2:1}.
Using the above conditions iteratively one can find all stability conditions computed for all 2-field and 
3-field directions as given in \cite{Arhrib:2011uy_vs}. 
 
We find the stability criteria for  all 2-field directions using copositivity and those match exactly with the similar 
conditions given by Ahrib $et.al.$ In \cite{Arhrib:2011uy_vs} the 3-field conditions are more involved and some time 
different set of conditions can be proposed depending upon which way one compose
the multiple fields. But in the method of finding {\sc cop} criteria the symmetric matrices out of the  quadratic 
forms are more fundamental. Thus the conditions are simple, precise and unique.  Moreover, invariance of copositive 
criteria under permutation ensures the unique set of conditions for a given quadratic form $x^T \Lambda x \geq 0$. 
We have constructed all 2- and 3-field directional potentials and computed respective copositivity conditions in 
appendix \ref{app:vac_Type_II} for interested readers. Here we discuss some of the interesting observation as mentioned above.
 
As our first example we have chosen a simpler 3-field direction that contains the field directions 
$\phi^0\,,\phi^+\; \textrm{and}\;\delta^+$ and also noted down as potential term in eqn.~\ref{3FV10_tII}. 
Using the above mentioned method suggested in \cite{Arhrib:2011uy_vs} one can calculate the necessary stability conditions as
\bea \label{eq:ahrib1}
\la > 0 \;\;\Band\;\; \la_2+\frac{\la_3}{2} > 0 \;\;\Band\;\; 2\la_1+\la_4 + \sqrt{2\la(2 \la_2+\la_3)} > 0 \;\Band\; \nonumber  \\ 
\left(2\la_1+\la_4 > 0\;\; \Bor \;\;  2\la(2\la_2+\la_3) > (2\la_1+\la_4)^2 \right).
\eea
These expressions exactly match with criteria given in \cite{Arhrib:2011uy_vs}. 
For the same potential, the independent {\sc cop} criteria\footnote{The fourth term in eqn.~\ref{3FV10_tII:1} can be simplified and written in terms of other conditions.} as shown in  eqn.~\ref{3FV10_tII:1}, can be noted in compact form as 
\be
\la > 0 \;\;\Band\;\; \la_2+\frac{\la_3}{2} > 0 \;\;\Band\;\; 2\la_1+\la_4 + \sqrt{2\la(2 \la_2+\la_3)} > 0. 
\ee
Here both the methods give the same allowed parameter space since the additional part in eqn.~\ref{eq:ahrib1} does not put any new constraint. 

In a second example we consider two 3-field directions ($\phi^0,\,\delta^+,\,\delta^0$) and ($\phi^+,\,\delta^+,\,\delta^{++}$). 
Corresponding stability conditions are more complicated and given in eqns.~B.27 and B.33 in \cite{Arhrib:2011uy_vs}.
Looking throughout the parameter space we have verified that these different looking conditions in fact cover {\sl the same parameter space}. 
Instead of two apparently different looking conditions, {\sl invariance of copositivity ensures a single set of stability conditions 
for both the directions}  $^{3F}V_2$ and $^{3F}V_3$ as presented in eqns.~\ref{3FV2_tII} and~\ref{3FV3_tII}. This is ensured by the following property of copositivity: invariance under permutations. 
Note that basis are different 
in each case, however that is not problem since we are only interested to find the conditions for stability in these situations, 
and as expected the basis dependency does not matter at all.
The stability conditions calculated using copositivity ensure the boundedness of the potential with mathematical confirmations. Moreover it
{\sl ensures the minimum allowed conditions resulting into more parameter space} compared to the conditions obtained 
from the successive squaring method. This was also verified numerically that the {\sc cop} conditions correspond to $^{3F}V_2$ (or, $^{3F}V_3$)
indeed allow more parameter space. In several cases these features are repeated and readers can easily identify them from the detailed list 
presented in appendix~\ref{app:vac_Type_II_3field}.


 \subsection{LR Model with Triplet Scalars }\label{sec:vac_LR_triplet}

We have adopted the model as suggested in \cite{Gunion:1989in_LRHiggs,Deshpande:1990ip_LRT,Rothstein:1990qx_LRT} where one can find the details. The gauge symmetry, $SU(3)_C \otimes SU(2)_L\otimes SU(2)_R \otimes U(1)_{B-L}$, is same as discussed in section~\ref{sec:vac_LR_doublet}.
In this model parity is spontaneously violated \cite{spont_parity_1,spont_parity_2,spont_parity_3,spont_parity_4}. Neutrino masses can be generated through Type-I and Type-II seesaw mechanisms. 
The model under consideration consists of a bi-doublet ($\Phi$) and two triplet scalars ($\Delta_{L/R}$). 
The component fields of these scalars can be expressed as following:
\begin{center}
 $
\hskip 25pt
\Phi =\left( \begin{array}{lr}
         \phi_1^0 \;&\; \phi_1^+\\ \\
         \phi_2^- & \phi_2^0
        \end{array} 
        \right)  \hskip 10pt , \hskip 10pt       
\Delta_{L,R}  = \left( \begin{array}{cc}
         \delta_{L,R}^+/\sqrt{2} & \delta_{L,R}^{++} \\  \\      
         \delta_{L,R}^0 & -\delta_{L,R}^+/\sqrt{2}
        \end{array} 
        \right) .
$
\end{center}

Once the neutral components of these scalars acquire vacuum expectation values the Left-Right as well as electroweak symmetries are broken spontaneously. 
Vacuum expectation values of these  scalars  can be written as:
\be\label{LRT_vev_struct} 
\left< \Phi \right>  =\left( \begin{array}{cc}
                        v_1 & 0\\
                        0      \;&\; v_2
                       \end{array}\right) , \hskip 20 pt                       
\left< \Delta_L \right>  =\left( \begin{array}{lr}
                        0   & 0\\
                        v_L\; & \;0
                       \end{array}\right) , \hskip 20 pt                       
\left< \Delta_R \right>  = \left(\begin{array}{lr}
                        0   & 0\\
                        v_R \;&\; 0
                       \end{array} \right),
\ee
where, for simplicity we set $v_2, v_L = 0$ without loss of generality. 

The scalar potential is given in appendix~\ref{app:vac_LR_triplet}. 
In the scalar potential there are total fifteen\footnote{Some of them are not appearing in our analysis as we have set $v_L=0$.} independent quartic couplings but only five out of them,  
$\la_1,\la_5,\la_6,\la_7\;\; \textrm{and}\;\; \la_{12}$, are related to the dominantly contributing terms (proportional to the $SU(2)_R\otimes U(1)_{B-L}$ breaking scale $\sim v_R$) in the scalar masses.
The other quartic couplings, $\la_2, \la_3, \la_4, \la_8, \la_9, \la_{10}, \la_{11}$, are involved with the contributions proportional to the $vev$ of the bi-doublet scalar, $v_1$. As $v_R>>v_1$ the later contributions are very small. We have treated all the heavy scalars to be almost degenerate by neglecting the contributions proportional to $v_1^2$. 
Here we have considered
$\la_9 \geq 0$\footnote{Some of the quartic couplings cannot be readily reconstructed from the masses of the scalars thus they are free parameters 
in our scenario. We have defined them as universal parameters which are degenerate \cite{JD-PK-TM}.} 
and $\la_{12}$ to be positive to get rid of the tachyonic scalars. This model is extensively described in several works we referred earlier. So we will directly move to the point to show the procedure of constructing the respective quadratic forms for different field directions and  thereafter calculate the necessary {\sc cop} conditions corresponding those. We have listed 
respective copositivity conditions in appendix~\ref{app:vac_LR_triplet_COP} for the sake of completeness.

While computing the {\sc cop} conditions we have also encountered the similar situations as in 
section~\ref{sec:vac_LR_doublet} and dealt them with same spirit.
We consider one such example where single symmetric matrix coming from two different directions $^{3F}V_6$ and $^{3F}V_7$. 
Corresponding quadratic forms are followed in  eqns.~\ref{eq:3FV6_1} or \ref{eq:3FV6_2}. 
Similar to our discussion at the previous section~\ref{sec:vac_LR_triplet}, we encounter one such unphysical 
parameter $C$. Here we have also illustrated the removal of superficial parameters from the final vacuum stability conditions.
Theoretically, as it is stated earlier, in the $\lambda_i$ parameter space  one needs to vary the parameter $C$ 
for all possible range (which is $[-\infty,\infty]$) together with conditions depicted in eqn.~\ref{eq:3FV6_c} as well as 
in eqns.~\ref{eq:3FV6_c1},~\ref{eq:3FV6_c2},~\ref{eq:3FV6_c3} (which are once again listed as following) and take union of all such allowed regions to achieve the full parameter space.
We have categorised the $C$ dependency as following:
\begin{enumerate}[(a)]
 \item \underline{ $ C  \geq  0 \;\mbox{such that} \; (1-C) \geq 0,$ i.e.,  $C \in [0,1]$:} 
 \bea  \label{eq:3FV6_c1_main}
\la_5+2\,\la_6 \geq 0, 
\eea

 \item \underline{ $C  \geq  0 \;\mbox{such that} \; (1-C) \leq 0,$ i.e., $C \in (1,\infty]$:}
 \bea  \label{eq:3FV6_c2_main}
\la_5+2\,\la_6 \geq 0  \hskip 0.8cm \Band \hskip 0.8cm  \la_5^2 - (1-C)^2 (\la_5+2\,\la_6)^2 \geq 0,
\eea

\item \underline{ $C < 0\;\mbox{such that} \; (1-C) \geq 0,$ i.e., $C \in [-\infty,0)$:}
\bea  \label{eq:3FV6_c3_main}
\la_5+2\,\la_6 \leq 0  \hskip 0.8cm \Band \hskip 0.8cm  \la_5^2 - (1-C)^2 (\la_5+2\,\la_6)^2 \geq 0.
\eea
\end{enumerate}
However, this procedure is highly impractical to implement in real calculations. 
But, as earlier, it is possible to remove the presence of $C$ from the final copositive conditions 
through the careful inspection of all the copositivity criteria. By combining last two cases (b and c) 
we can write $\la_5 \geq |(1-C)(\la_5+2\la_6)|$, with $C \in [-\infty,\infty]$ excluding the range $[0,1]$. 
It is obvious that the other derived condition $\la_5 \geq 0$ is more relaxed for the given range of $C$. 
Thus $\la_5 \geq 0$ allows larger parameter space than the condition $\la_5 \geq |(1-C)(\la_5+2\la_6)|$.
Now the case (a)  posses the criteria $\la_5+2\la_6 \geq 0$ that leads to the less parameter space than already derived condition $\la_5+\la_6 \geq 0$ for $\la_6 <0$. 
For $\la_6 \geq 0$ both conditions are automatically satisfied as $\la_5 \geq 0$.

\begin{figure}[tbh]
\begin{center}
\includegraphics[width=6cm,angle=0]{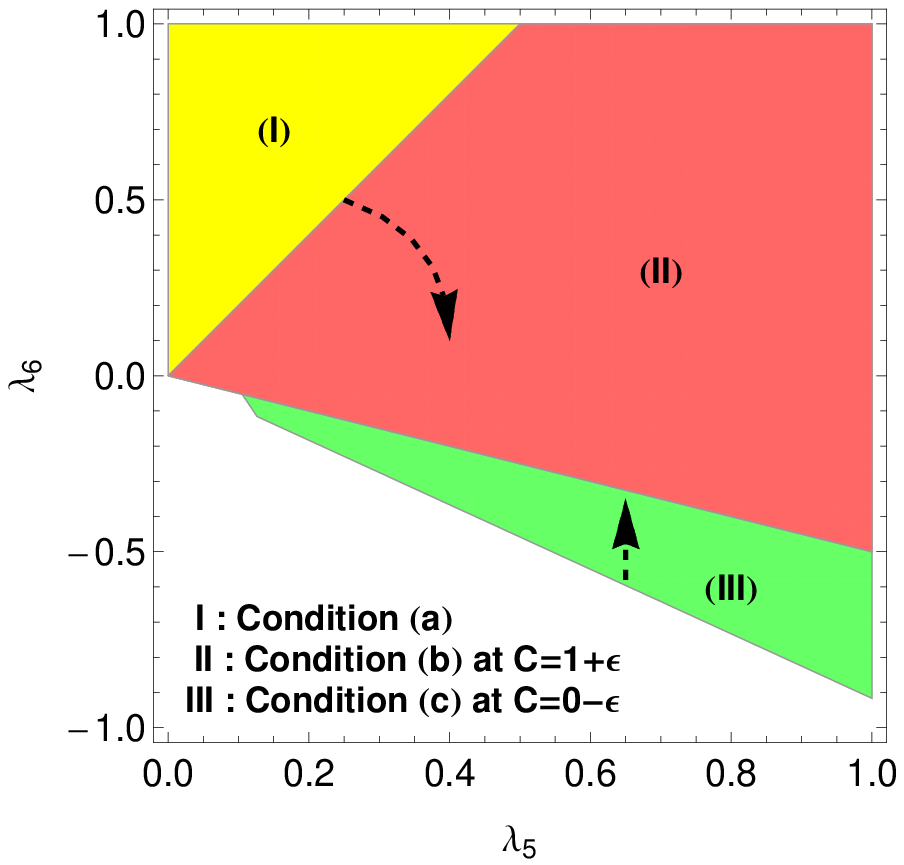}
\hspace{1cm}
\includegraphics[width=6cm,angle=0]{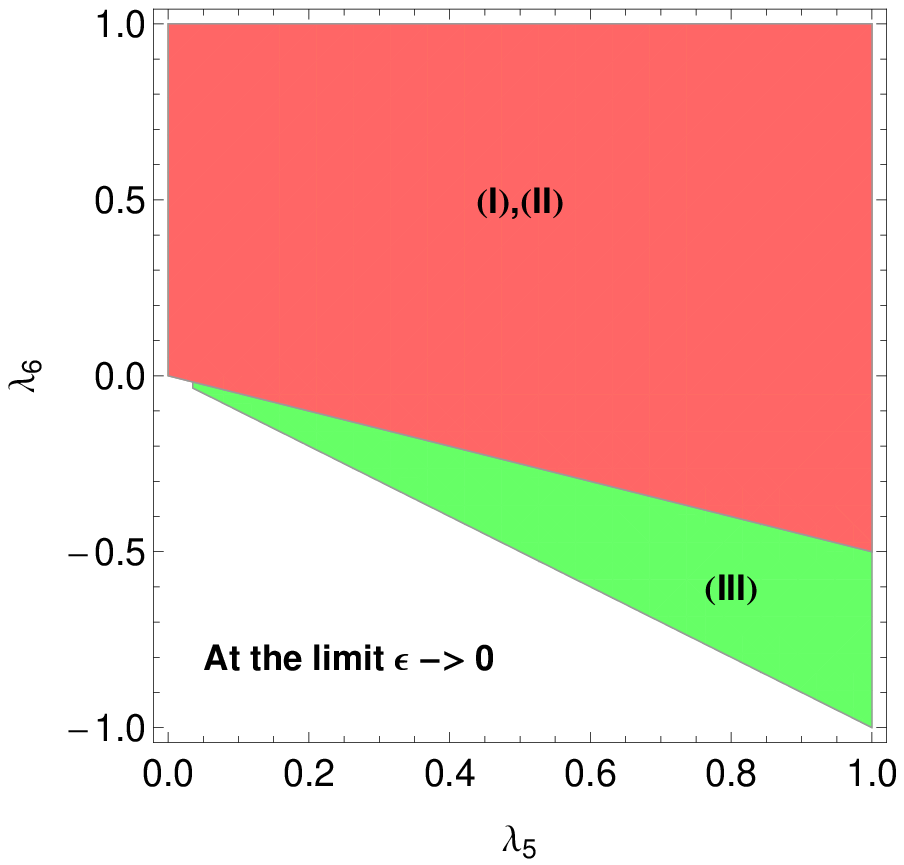}
\caption{Maximizing allowed parameter space in $\lambda_5-\lambda_6$ plane correspond to suitable parameter $C$ in the case of $^{3F}V_6$ and $^{3F}V_7$.
Figure~(a): small value of $\epsilon$ (= 0.2) chosen to parametrise $C$. Black arrows represents the movements of boundary lines for increased values of $\epsilon$ satisfying the conditions.
This is to demonstrate that as the limit $\epsilon$ goes to zero, maximum allowed space is achieved as shown in Figure~(b). 
Clearly these set of conditions at the $C$ values which maximise the allowed parameters, 
equivalent to a simple condition given by $(\la_5+\la_6) \geq 0$ which is independent of $C$.}
 \label{fig:3FV6_3FV7}
 \end{center}
 \end{figure}

This can also be demonstrated pictorially in the following manner.
Assuming that we are interested in the region of $|\lambda_i| \leq 1$ from perturbativity, 
we would like to find out particular values of $C$ which actually maximise 
the  allowed regions for a given condition. Thereafter simply rewrite those conditions at that 
point and demand that to be the final condition.  At our present example
conditions given in eqns.~\ref{eq:3FV6_c2_main} and  \ref{eq:3FV6_c3_main} depend on the parameter $C$. However, as 
demonstrated in Figure~(\ref{fig:3FV6_3FV7}) they maximise the allowed parameters at $C=1$ and $C=0$ respectively. 
Moreover, union of these set of nontrivial conditions for different ranges of $C$ equivalent to a simple condition 
given by $(\la_5+\la_6) \geq 0$ which is independent of $C$. Interestingly, this is not a new condition and already
explored in several 2-field copositivity conditions like $^{2F}V_8$ and $^{2F}V_{10}$. Final copositivity conditions in
this case can be written as \ref{eq:3FV6_final}.
Including all copositive conditions we get a final set of conditions as,
$$\la_1 \geq 0 \;\;\Band\;\; \la_5 \geq 0 \;\;\Band\;\; \la_5+\la_6 \geq 0 \;\;\Band\;\;  16\;\la_1\,\la_5 - \la_{12}^2 \geq 0.$$

\section{Copositivity using Principal Sub-matrices}\label{sec:COP-PS}
  
In the earlier section we have shown how one can map the part of the scalar potential that contains only quartic couplings 
to the quadratic form and define the symmetric matrices. Then using the proposal for order two, three, and four 
matrices one can compute the copositive criteria, i.e., the conditions for vacuum stability. But 
this analysis can be used for generic symmetric matrices of order up to four and for some special higher order matrices too.  
To avoid such restrictions one can use the principal sub-matrix formalism in much broader sense to check the copositivity of symmetric matrices. This is not restricted 
up to any particular order of matrices. The sole idea of copositivity can be understood in terms of the eigenforms 
(eigenvalues and respective eigenvectors) of the principal sub-matrices.

To understand this idea of copositive criteria here we examine with an explicit example.
Let us consider a quadratic form:\\
$$ x^T \Lambda_1 x = x_1^2 +2x_2^2 + 3x_3^2 -2x_1 x_2 -2 x_1 x_3 -4 x_2 x_3, $$
and the corresponding symmetric matrix  of order three\footnote{As for symmetric matrices $(\Lambda)_{ij}=(\Lambda)_{ji}$, 
we are not writing the full matrix. The upper-triangular matrix with the diagonal elements are sufficient to represent the 
full matrix. Through out the text we have used this notation to represent the symmetric matrices.} in the basis
of $\{x_1,x_2,x_3\}$ can be given as:
\be \label{eq:examp_1}
\Lambda_1 = \left( 
		\begin{array}{*7{C{1.5cm}}}
                1  & -1 & -1\\\cline{1-1}
                \bord & 2 & -2 \\\cline{2-2}
                 & \bord & 3\\\cline{3-3}
               \end{array}
               \right).
\ee
This matrix has seven\footnote{Any matrix of order $n$ has $(2^n -1)$ principal sub-matrices.} principal sub-matrices:  three of order one, three of order two, and one of order three.
The principal sub-matrices of order one, i.e., the diagonal elements are positive. The eigenvalues of order two principal sub-matrices are all 
positive.
But order three principal sub-matrix has a negative eigenvalue that corresponds to a positive eigenvector thus $\Lambda_1$ is not copositive.

Let us consider another quadratic form:\\
$$ x^T \Lambda_2 x = 2x_1^2 +2x_2^2 + x_3^2 -2x_1 x_2 -2 x_1 x_3 + 4 x_2 x_3,$$ 
and similarly the corresponding symmetric matrix of order three is given as
\be \label{eq:examp_2}
\Lambda_2 = \left( 
		\begin{array}{*7{C{1.5cm}}}
                2  & -1 & -1\\ \cline{1-1}
                \bord & 2 & 2 \\ \cline{2-2}
                 & \bord & 1\\ \cline{3-3}
               \end{array}
               \right).
\ee
Here we find one of the order two principal sub-matrices and order three principal sub-matrix have negative eigenvalues. 
But unlike the previous case here the negative eigenvalues are associated with negative eigenvectors. Thus this matrix, $\Lambda_2$, is copositive.

This procedure is useful while we are dealing with a matrix numerically. While adjudging the validity of a model up to a certain scale we need to perform the renormalisation group evolutions of the quartic couplings belong to the scalar potential. There this matrix can be constructed out of these quartic couplings at each scale and one can check the copositivity using this method. We have provided a brief algorithm encoded in mathematica for this approach in appendix~\ref{app:Algo-PS}.

 \section{Conclusion}
 
The vacuum stability is an important issue that must be addressed for any beyond standard model scenario. 
Thus one needs to carefully examine the vacuum stability criteria that lead to the boundedness of the full 
scalar potential. In this paper we have revisited few phenomenologically interesting scenarios, like Left-Right 
symmetry with doublet, and triplet scalars and Type-II seesaw models to address and adjudge the stability of the vacuum. 
We have adopted a technique which is well discussed in the context of `Linear Algebra', 
namely {\it copositivity of symmetric matrix}. Here we have discussed two approaches to check the copositivity -- using the explicit structure of that matrix and other one using the principal sub-matrices. We have first discussed how to reconstruct 
the symmetric matrices using the quartic couplings and then compute the copositivity criteria to deduce the vacuum stability
conditions. We have performed a detailed and complete analysis for Left-Right symmetric model with doublet scalars, 
and compute the full set of vacuum stability criteria. Then we have performed a comparative study 
among this procedure and earlier used method (as suggested by Ahrib $et.al.$) and shown the advantages of our method. 
We have then used this method for much more complicated models like Left-Right symmetry with triplet scalars. 
We have computed the stability criteria for these models for different field directions. Apart from this analytical but in 
some sense restricted procedure we also discuss an alternative method to check whether a matrix is copositive or not using principal 
sub-matrix formalism. For the principal sub-matrix approach we have provided an general algorithm 
to check the copositivity of a symmetric matrix of finite order and also provide explicit numerical example.

\begin{acknowledgments}
Authors would like to acknowledge Soumya Das, Amitava Raychaudhuri, and R. Thangadurai for useful discussions. Work of JC is supported by Department 
of Science and Technology, Government of INDIA under the Grant Agreement number IFA12-PH-34 (INSPIRE Faculty Award).
 TM acknowledges the hospitality of Indian Institute of Technology, Kanpur where part of this work was completed.
\end{acknowledgments}

\appendix


\section{Copositivity ({\sc cop}) Conditions} \label{app:copositive_conditions}

For a general case where the diagonal and off-diagonal elements are free parameters it is difficult to check whether the matrix 
is copositive or not. Thus here we mention the explicit criteria of the copositivity for the matrices of order two, three and four.

\subsection{Copositivity of order two matrix} \label{app:cop-order-2}
 Let us consider a symmetric matrix of order two:
\be \label{eq:copositive-matrix-order-2}
\mathcal{S}_2 = \left( 
		\begin{array}{*7{C{1.5cm}}}
                \la_{11}  & \la_{12}\\\cline{1-1}
                \bord & \la_{22}\\\cline{2-2}
               \end{array}
               \right).
\ee
This matrix is copositive if and only if 
\be \label{eq:copositive-matrix-order-2:1}
\la_{11}\geq 0,\;\;\; \la_{22}\geq 0,\;\;\;\textrm{and}\;\;\; \la_{12}+ \sqrt{\la_{11}\la_{22}}\geq 0.
\ee

\subsection{Copositivity of order three matrix} \label{app:cop-order-3}
Let us consider a symmetric matrix of order three:
\be \label{eq:copositive-matrix-order-3}
\mathcal{S}_3 = \left( 
		\begin{array}{*7{C{1.5cm}}}
                \la_{11}  & \la_{12} & \la_{13}\\ \cline{1-1}
                \bord & \la_{22} & \la_{23}\\ \cline{2-2}
                 & \bord & \la_{33}\\ \cline{3-3}
               \end{array}
               \right).
\ee
This matrix copositive if and only if,
\begin{align} \nonumber
& \la_{ii}\geq 0,\;\;\;
 \la_{ij}+\sqrt{\la_{ii}\la_{jj}}  \geq 0, \\
& \sqrt{\prod_{i=1,2,3}\la_{ii}}+\sum_{i,j,k}\la_{ij}\sqrt{\la_{kk}}+\sqrt{2\,\prod_{i,j,k}(\la_{ij}+\sqrt{\la_{ii}\la_{jj}})} \;\; \geq  0 ,
\end{align}
where $\{i,j,k\}$ = Permutation of $\{1,2,3\}$ with $i<j$.

\subsection{Copositivity of order four matrix} \label{app:cop-order-4}
Let us consider a symmetric matrix of order four:
\be \label{eq:copositive-matrix-order-4}
\mathcal{S}_4 = \left( 
		\begin{array}{*7{C{1.5cm}}}
                \la_{11}  & \la_{12} & \la_{13} & \la_{14}\\ \cline{1-1}
                \bord & \la_{22} & \la_{23} & \la_{24}\\ \cline{2-2}
                 & \bord & \la_{33} & \la_{34}\\ \cline{3-3}
                 &  & \bord & \la_{44}\\ \cline{4-4}
               \end{array}
               \right).
\ee
To determine whether this matrix is copositive or not we need to adjudge eight different cases depending on the sign distributions of the off-diagonal elements. 
But in all cases one generic condition, all the diagonal elements are positive, must be satisfied. 
We have successively discussed them as suggested in \cite{Li_Feng}. From here onwards $\{i,j,k,l\}$ are any permutation of $\{1,2,3,4\}$ and  $i\neq j \neq k \neq l$.
\begin{enumerate}[(I)]
 \item[Case I:] If all the off-diagonal elements of $\mathcal{S}_4$ are positive then this is copositive if and only if $\la_{ii}\geq0$.
  \item[Case II:]If $\la_{ij}\leq 0$ and other off-diagonal elements are positive then $\mathcal{S}_4$ is copositive if and only if 
  $(\la_{ii}\la_{jj}-\la_{ij}^2) \geq 0$.
 \item[Case III:] If $\la_{ij},\la_{lk}\leq 0$ and other off-diagonal elements are positive then the matrix is copositive if and only if 
 $(\la_{ii}\la_{jj}-\la_{ij}^2) \geq 0$, $(\la_{ll}\la_{kk}-\la_{lk}^2) \geq 0$.
 \item[Case IV:] If $\la_{ij},\la_{ik}\leq 0$ then we must have 
 $\big(\la_{ii}\la_{jk}-\la_{ij}\la_{ik}+\sqrt{(\la_{ii}\la_{jj}-\la_{ij}^2) (\la_{ii}\la_{kk}-\la_{ik}^2)} ~\big) \geq 0$ 
 to make this matrix copositive.
 \item[Case V:]  If $\la_{ij},\la_{jk},\la_{ik}\leq 0$ while the other off-diagonal elements are positive then $\mathcal{S}_4$ is copositive 
 if and only if the following order three matrix is copositive:
\be \nonumber \left( 
		\begin{array}{*7{C{1.5cm}}}
                \la_{ii}  & \la_{ij} & \la_{ik}\\ \cline{1-1}
                \bord & \la_{jj} & \la_{jk}\\ \cline{2-2}
                 & \bord & \la_{kk}\\ \cline{3-3}
               \end{array}
               \right).
\ee               
\item[Case VI:] If $\la_{ij},\la_{ik},\la_{il}\leq 0$ and other off-diagonal elements are positive then the following matrix:
\be  \nonumber \left( 
		\begin{array}{*7{C{3cm}}}
                \la_{ii}\la_{jj}-\la_{ij}^2  & \la_{ii}\la_{jk}-\la_{ij}\la_{ik} & \la_{ii}\la_{jl}-\la_{ij}\la_{il}\\\cline{1-1}
                \bord & \la_{ii}\la_{kk}-\la_{ik}^2 & \la_{ii}\la_{kl}-\la_{ik}\la_{il}\\\cline{2-2}
                & \bord & \la_{ii}\la_{ll}-\la_{il}^2\\\cline{3-3}
               \end{array}
               \right)
\ee 
must be copositive in order to make $\mathcal{S}_4$ to be copositive.
\item[Case VII:] If $\la_{ij},\la_{jk},\la_{kl}\leq 0$ and other off-diagonal elements are positive then we need to construct a matrix of order three which  
has to be copositive and that will imply $\mathcal{S}_4$ is copositive. Let us consider a matrix $\mathcal{S}_3^{'}$:
\be  \nonumber \left( 
\begin{array}{*7{C{3.6cm}}}
\la_{kk} \big(\la_{jj}\la_{ik}^2-2\la_{ij}\la_{ik}\la_{jk}+ \la_{ii} \la_{jk}^2\big)  & \la_{kk} (\la_{jj}\la_{ik} - \la_{ij} \la_{jk}) & \la_{kk}(\la_{ik}\la_{jl} - \la_{jk} \la_{il})\\ \cline{1-1} 
 \bord & & \cline{2-2}
 \bord & \la_{jj}\la_{kk} - \la_{jk}^2 & \la_{kk}\la_{jl} - \la_{jk} \la_{kl}\\ \cline{2-2} 
  &\bord & & \cline{3-3}
  & \bord & \la_{kk}\la_{ll} - \la_{kl}^2\\ \cline{3-5}
\end{array}
\right).
\ee   

Thus $\mathcal{S}_4$ matrix will be copositive if and only if $\mathcal{S}_3^{'}$ is copositive.
\item[Case VIII:] If $\la_{ij},\la_{jk},\la_{kl}, \la_{il} \leq 0$ and other off-diagonal elements are positive then similar to the Case VII one nedds to reconstruct 
a matrix of order three which has to be copositive in order to make $\mathcal{S}_4$ copositive. That matrix of order three should be $\mathcal{S}_3^{''}$ :
\be  \nonumber \left( 
		\begin{array}{*7{C{3.6cm}}}
                \la_{ll} (\la_{ii}\la_{jl}^2-2\la_{ij}\la_{il}\la_{jl}+ \la_{jj} \la_{il}^2)  &  \la_{ll} (\la_{ii}\la_{jl} - \la_{ij} \la_{il}) & \la_{ll}(\la_{ik}\la_{jl} - \la_{il} \la_{jk})\\ \cline{1-1}
                \bord & & \cline{2-2}
                \bord & \la_{ii}\la_{ll} - \la_{il}^2 & \la_{ll}\la_{ik} - \la_{il} \la_{kl}\\ \cline{2-2}
                &\bord & & \cline{3-3}
                & \bord & \la_{kk}\la_{ll} - \la_{kl}^2\\ \cline{3-5}
               \end{array} 
               \right).
\ee   
\end{enumerate}


\section{Scalar Potential}\label{app:scalar_potential}


\subsection{LR model with Doublet Scalar: Quartic Couplings of the Potential}
\label{app:vac_LR_doublet}

Scalar potential for LR model with doublet scalars can be written as \cite{Holthausen:2011aa_smRGE}:
\beas\label{app:eq:LR_doublet_pot}
V_{LRD}^{Q} (\Phi,H_L,H_R) &=&  4 \la_1 \Big(\Tr[\Phi^\dagger\Phi]\Big)^2 +
4 \la_2 \Big(\Tr[\Phi^\dagger \tilde{\Phi}]+\Tr[\Phi\tilde{\Phi}^\dagger]\Big)^2 +
4 \la_3 \Big(\Tr[\Phi^\dagger \tilde{\Phi}]-\Tr[\Phi\tilde{\Phi}^\dagger]\Big)^2 \\
&& + \frac{\kappa_1}{2} \Big(H_L^\dagger H_L+H_R^\dagger H_R\Big)^2 +
\frac{\kappa_2}{2} \Big(H_L^\dagger H_L-H_R^\dagger H_R\Big)^2 \\
&& + \beta_1 \Big(\Tr[\Phi^\dagger \tilde{\Phi}]+\Tr[\Phi\tilde{\Phi}^\dagger]\Big)\Big(H_L^\dagger H_L+H_R^\dagger H_R\Big)\\
&& + f_1\Big(H_L^\dagger\big(\tilde{\Phi}\tilde{\Phi}^\dagger-\Phi \Phi^\dagger) H_L
-H_R^\dagger \big(\Phi^\dagger\Phi-\tilde{\Phi}^\dagger \tilde{\Phi} \big)H_R \Big).
\eeas

\subsection{Type-II Seesaw: Quartic Couplings of the Potential}
\label{app:vac_Type_II_seesaw}

Scalar potential for Type--II seesaw can be written as \cite{Arhrib:2011uy_vs}: 
\beas\label{app:eq:LR_doublet_pot}
V_{Type-II}^{Q} (\Phi,\Delta) &=&   \frac{\la}{4} \Big(\Phi^\dagger\Phi\Big)^2 +
 \la_1 \left(\Phi^\dagger\,\Phi\right)\left(\Tr[\Delta^\dagger\Delta]\right) +
  \la_2 \left(\Tr[\Delta^\dagger\Delta]\right)^2 + 
   \la_3 \Tr[\left(\Delta^\dagger\Delta\right)^2] \\ && + 
    \la_4 \left(\Phi^\dagger\,\Delta\,\Delta^\dagger\Phi\right).
\eeas


\subsection{LR model with Triplet Scalar: Quartic Couplings of the Potential}
\label{app:vac_LR_triplet}

The most generic scalar potential in this model can be written as given in~\cite{Gunion:1989in_LRHiggs,Deshpande:1990ip_LRT,Rothstein:1990qx_LRT}: 

\beas\label{eq:LR_triplet_pot}
 && V_{LRT}^{Q}(\Phi,\Delta_L,\Delta_R) = \\
    &+& \lambda_1\bigg\{\Big(\Tr\big[\Phi^\dagger \Phi\big]\Big)^2\bigg\} +
    \lambda_2\bigg\{ \Big(\Tr\big[\tilde{\Phi}\Phi^\dagger\big]\Big)^2+\Big(\Tr\big[\tilde{\Phi}^\dagger \Phi\big]\Big)^2 \bigg\} 
    + \lambda_3\bigg\{\Tr\big[\tilde{\Phi}\Phi^\dagger\big]\Tr\big[\tilde{\Phi}^\dagger \Phi\big] \bigg\}\\
    &+& \lambda_4 \bigg\{ \Tr\big[\Phi^\dagger \Phi\big]\Big(\Tr\big[\tilde{\Phi}\Phi^\dagger\big]
    +\Tr\big[\tilde{\Phi}^\dagger \Phi\big]\Big) \bigg\}
    +\lambda_5\bigg\{ \Big(\Tr\big[\Delta_L \Delta_L^\dagger\big]\Big)^2+\Big(\Delta_R \Delta_R^\dagger\Big)^2 \bigg\} \\
    &+& \lambda_6 \bigg\{\Tr\big[\Delta_L \Delta_L\big]\;\Tr\big[\Delta_L^\dagger \Delta_L^\dagger\big]
    +\Tr\big[\Delta_R \Delta_R\big]\;\Tr\big[\Delta_R^\dagger \Delta_R^\dagger\big]  \bigg\}
    + \lambda_7 \bigg\{\Tr\big[\Delta_L\Delta_L^\dagger\big]\;\Tr\big[\Delta_R\Delta_R^\dagger\big]\bigg\} \\
    &+& \lambda_8[\Delta_L\Delta_L^\dagger\big] \bigg\{\Tr\big[\Delta_L\Delta_L^\dagger\big]\;
    \Tr\big[\Delta_R\Delta_R^\dagger\big] \bigg\}
    + \lambda_9 \bigg\{\Tr\big[\Phi^\dagger \Phi\big]\Big(\Tr\big[\Delta_L\Delta_L^\dagger\big]
    +\Tr\big[\Delta_R\Delta_R^\dagger\big]\Big)\bigg\} \\
    &+& (\lambda_{10}+i\,\lambda_{11}) 
    \bigg\{\Tr\big[\Phi\tilde{\Phi}^\dagger\big]\Tr\big[\Delta_R\Delta_R^\dagger\big] 
    + \Tr\big[\Phi^\dagger\tilde{\Phi}\big]\Tr\big[\Delta_L\Delta_L^\dagger\big]\bigg\} \\
    &+& (\lambda_{10}-i\,\lambda_{11}) 
    \bigg\{\Tr\big[\Phi^\dagger\tilde{\Phi}\big]\Tr\big[\Delta_R\Delta_R^\dagger\big] 
    + \Tr\big[\tilde{\Phi}^\dagger\Phi\big]\Tr\big[\Delta_L\Delta_L^\dagger\big]\bigg\} \\
    &+& \lambda_{12} \bigg\{ \Tr\big[\Phi \Phi^\dagger \Delta_L \Delta_L^\dagger\big]
    +\Tr\big[\Phi^\dagger \Phi \Delta_R \Delta_R^\dagger\big]\bigg\}
    + \lambda_{13} \bigg\{\Tr\big[\Phi\Delta_R\Phi^\dagger\Delta_L^\dagger\big]
    +\Tr\big[\Phi^\dagger\Delta_L\Phi\Delta_R^\dagger\big] \bigg\} \\
    &+& \lambda_{14}\bigg\{\Tr\big[\tilde{\Phi}\Delta_R\Phi^\dagger\Delta_L^\dagger\big]
    +\Tr\big[\tilde{\Phi}^\dagger\Delta_L\Phi\Delta_R^\dagger\big]\bigg\} 
    + \lambda_{15} \bigg\{\Tr\big[\Phi\Delta_R\tilde{\Phi}^\dagger\Delta_L^\dagger\big]
    +\Tr\big[\Phi^\dagger\Delta_L\tilde{\Phi}\Delta_R^\dagger\big]\bigg\},
\eeas
where all the coupling constants are real. 



\section{Conditions of {\sc cop}: LR Model with Doublet Scalars}\label{app:vac_LR_doublet_COP}

\subsection{2-Field Directions and Stability Conditions}\label{sec:vac_LR_doublet_2field}

\begin{enumerate}[(I)]

\item
\be
^{2F}V_1(\phi_1^0\,,\,\phi_1^+)  = \la_1\;\left({\phi_1^0}^2+{\phi_1^+}^2\right)^2. 
\ee
Here we would like to note that all the field directions are evaluated in terms of the modulus of each field, i.e., $\phi_1^0 \equiv |\phi_1^0|$, and we have used the same notation through out the literature. 
 
In matrix form it can be represented in basis $({\phi_1^0}^2,{\phi_1^+}^2)$:

$$
\left(
\begin{array}{*3{C{1.5cm}}}
\la_1 & \la_1\\ \cline{1-1}
\bord & \la_1 \\ \cline{2-2}
\end{array}
\right).
$$

\underline{Copositivity  condition}: \\
$$ \la_1  \geq  0. $$

\item
\be
 ^{2F}V_2(\phi_1^+\,,\,h_R^+) =  \la_1\;{\phi_1^+}^4 + \frac{2\beta_1+f_1}{2}{h_R^+}^2 {\phi_1^+}^2.
\ee

\be
^{2F}V_3(\phi_1^0\,,\,h_R^0) = \la_1\;{\phi_1^0}^4 + \frac{2\beta_1+f_1}{2}{h_R^0}^2 {\phi_1^0}^2 .
\ee

In matrix form each of them can be represented in basis $({\phi_1^+}^2 \Leftrightarrow {\phi_1^0}^2, {h_R^+}^2 \Leftrightarrow {h_R^0}^2 )$:

$$
\left(
\begin{array}{*3{C{1.5cm}}}
\la_1 & \frac{2\beta_1+f_1}{4}\\ \cline{1-1}
\bord & 0 \\ \cline{2-2}
\end{array}
\right).
$$
Here two different quadratic forms are represented by the same matrix in different basis and the $sign$ `$\Leftrightarrow$' implies the mutual exchange of fields leading one quadratic form to other one.
For example, if we replace ${\phi_1^+}^2$ and ${h_R^+}^2$ in $^{2F}V_2(\phi_1^+\,,\,h_R^+)$ by ${\phi_1^0}^2$ and ${h_R^0}^2$ simultaneously then we achieve $^{2F}V_3(\phi_1^0\,,\,h_R^0)$.
We have used the same notation through out the text.

\underline{Copositivity  condition}: \\
$$ \la_1 \geq 0, \hskip 1cm  2\beta_1+f_1 \geq  0. $$

\item
\be
 ^{2F}V_4(\phi_1^0\,,\,h_R^+)  = \la_1\;{\phi_1^0}^4 + \frac{2\beta_1-f_1}{2}{h_R^+}^2 {\phi_1^0}^2. 
\ee

\be
^{2F}V_5(\phi_1^+\,,\,h_R^0)   = \la_1\;{\phi_1^+}^4 + \frac{2\beta_1-f_1}{2}{h_R^0}^2 {\phi_1^+}^2. 
\ee

In matrix form both of them can be represented in basis $({\phi_1^0}^2 \Leftrightarrow {\phi_1^+}^2, {h_R^+}^2 \Leftrightarrow {h_R^0}^2 )$:

$$
\left(
\begin{array}{*3{C{1.5cm}}}
\la_1 & \frac{2\beta_1-f_1}{4}\\ \cline{1-1}
\bord & 0 \\ \cline{2-2}
\end{array}
\right).
$$

\underline{Copositivity  condition}: \\
$$ \la_1  \geq  0, \hskip 1cm   2\beta_1-f_1 \geq  0. $$

\end{enumerate}


\subsection{3-Field Directions and Stability Conditions}\label{sec:vac_LR_doublet_3field}

\begin{enumerate}[(I)]
\item
\bea
^{3F}V_1 (\phi_1^0\,,\,\phi_1^+\,,\,h_R^{0})  &=&  {h_R^0}^2\left(\beta_1({\phi_1^0}^2+{\phi_1^+}^2)+\frac{1}{2}f_1({\phi_1^0}^2-{\phi_1^+}^2)\right) \nonumber \\
&+& \la_1 \left({\phi_1^0}^2+{\phi_1^+}^2\right)^2. 
\eea

\bea
^{3F}V_2 (\phi_1^0\,,\,\phi_1^+\,,\,h_R^{+})  &=&   {h_R^+}^2\left(\beta_1({\phi_1^0}^2+{\phi_1^+}^2)+\frac{1}{2}f_1({\phi_1^+}^2-{\phi_1^0}^2)\right) \nonumber\\
& + &\la_1 \left({\phi_1^0}^2+{\phi_1^+}^2\right)^2.
\eea

In matrix form each can be represented in basis $({\phi_1^0}^2, {\phi_1^+}^2, {h_R^0}^2 \Leftrightarrow {h_R^+}^2 )$:

$$
\left(
\begin{array}{*4{C{1.5cm}}}
\la_1 & \la_1 & \frac{2\beta_1-f_1}{4}\\ \cline{1-1}
\bord & \la_1 & \frac{2\beta_1+f_1}{4} \\ \cline{2-2}
& \bord & 0 \\ \cline{3-3}
\end{array}
\right).
$$

\underline{Copositivity conditions}: \\
$$ \la_1  \geq  0, \hskip 0.8cm 2\beta_1-f_1  \geq 0, \hskip 0.8cm 2\beta_1+f_1  \geq  0. $$

\item
\be
^{3F}V_3 (\phi_1^0\,,\,h_R^{0}\,,\,h_R^{+})  = \frac{1}{2}{\phi_1^0}^2\bigg(f_1\left({h_R^0}^2-{h_R^+}^2\right)
                                                   +2\beta_1\left({h_R^0}^2+{h_R^+}^2\right)+2\la_1{\phi_1^0}^2\bigg).
\ee

\be
^{3F}V_4 (\phi_1^+\,,\,h_R^{0}\,,\,h_R^{+})  = \frac{1}{2}{\phi_1^+}^2\bigg(f_1\left({h_R^+}^2-{h_R^0}^2\right)
                                                  +2\beta_1\left({h_R^0}^2+{h_R^+}^2\right)+2\la_1{\phi_1^+}^2\bigg).
\ee

In matrix form both can be represented in basis $({\phi_1^0}^2 \Leftrightarrow {\phi_1^+}^2, {h_R^0}^2, {h_R^+}^2)$:

$$
\left(
\begin{array}{*4{C{1.5cm}}}
\la_1 & \frac{2\beta_1+f_1}{4} & \frac{2\beta_1-f_1}{4}\\ \cline{1-1}
\bord & 0 & 0  \\ \cline{2-2}
&\bord & 0 \\ \cline{3-3}
\end{array}
\right).
$$

\underline{Copositivity  condition}: \\
$$ \la_1  \geq  0, \hskip 0.8cm 2\beta_1-f_1  \geq 0, \hskip 0.8cm 2\beta_1+f_1  \geq  0. $$

\end{enumerate}


\subsection{4-Field Directions and Stability Conditions}\label{sec:vac_LR_doublet_4field}

\begin{enumerate}[(I)]
\item
 \bea \label{eq:vac_LR_doublet_4field}
^{4F}V_1 (\phi_1^0,\phi_1^+,h_R^{0},h_R^{+})&=&\frac{1}{2}\Bigg(f_1\left(h_R^+(\phi_1^0-\phi_1^+)+h_R^0(\phi_1^0+\phi_1^+)\right)
                         \left(h_R^0(\phi_1^0-\phi_1^+) \nonumber  - h_R^+(\phi_1^0+\phi_1^+)\right)   \\ & &  +2({\phi_1^0}^2+{\phi_1^+}^2) \left(({h_R^0}^2+{h_R^+}^2)\beta_1+\la_1({\phi_1^0}^2+{\phi_1^+}^2)\right)\Bigg).
\eea

 In matrix form it can be represented in basis $({\phi_1^0}^2, {\phi_1^+}^2, {h_R^{0}}^2, {h_R^{+}}^2, \phi_1^0 \phi_1^+, h_R^0 h_R^+)$:
$$
\left(
\begin{array}{*7{C{2.2cm}}}
\la_1 & \la_1 & \frac{(1-C)}{2} \frac{f_1+2\beta_1}{2} & \frac{(1-K)}{2} \frac{2\beta_1-f_1}{2} & 0 & 0 \\ \cline{1-1}
\bord & \la_1 &  \frac{2\beta_1-f_1}{4}              & \frac{f_1+2\beta_1}{4}               & 0 & 0 \\ \cline{2-2}
      & \bord & 0        & 0        & 0 & 0 \\ \cline{3-3}
      &       & \bord    & 0        & 0& 0 \\ \cline{4-4}
      &       &          & \bord    & C \frac{f_1+2\beta_1}{2} & -f_1 \\ \cline{5-5} 
      &       &          &          & \bord    & K \frac{f_1+2\beta_1}{2}\\ \cline{6-6}
\end{array}
\right).
$$
\underline{ Copositivity  conditions : } \\
$$\la_1  \geq  0,  \hskip 0.8cm  C (2\beta_1+f_1)  \geq  0, \hskip 0.8cm K (2\beta_1+f_1)   \geq  0, $$
$$ \hskip 0.8cm C\;K\,\left(\frac{2\beta_1+f_1}{2}\right)^2- f_1^2  \geq  0.  $$

In this case we do not have any other four field directions. Thus we can find the conditions which ensure that the potential is bounded from below by combining all {\sc cop} criteria. From 2- and 3-fields directions we find the following conditions: $\la_1 \geq 0, \beta_1 \geq |f_1|/2$. 
We can eliminate both the unphysical parameters $C$ and $K$ from the {\sc cop} emerged from 4-field direction by demanding the maximisation of the parameter space. Detailed discussion is in 
section~\ref{sec:vac_LR_doublet} which leads to a condition we already have from 2- and 3-field directions.

\end{enumerate}

 %
  \section{Conditions of {\sc cop}: Type--II SeeSaw}\label{app:vac_Type_II}

\subsection{2-Field Directions and Stability Conditions}\label{app:vac_Type_II_2field}

\begin{enumerate}[(I)]
\item 
\be
^{2F}V_1(\phi^0\,,\,\phi^+)  = \frac{\la}{4}\;\left({\phi^0}^2+{\phi^+}^2\right)^2 .
\ee

In matrix form both of them can be represented in basis $({\phi^0}^2, {\phi^+}^2)$:

$$
\left(
\begin{array}{*3{C{1.5cm}}}
\la/4 & \la\\ \cline{1-1} 
\bord & \la/4 \\ \cline{2-2}
\end{array}
\right).
$$

\underline{Copositivity  condition}: \\
$$ \la \geq 0. $$

\item

\begin{align}
^{2F}V_2(\phi^+\,,\,\delta^{++}) & = (\la_2+\la_3) {\delta^{++}}^4 + (\la_1+\la_4) {\delta^{++}}^2\,{\phi^+}^2 + \frac{\la}{4} {\phi^+}^4. \\ \nonumber \\
^{2F}V_3(\phi^0\,,\,\delta^{0})  & = (\la_2+\la_3) {\delta^{0}}^4 + (\la_1+\la_4) {\phi^0}^2 {\delta^{0}}^2 + \frac{\la}{4} {\phi^0}^4.
\end{align}

In matrix form both can be represented in basis $({\phi^+}^2 \Leftrightarrow {\phi^0}^2, {\delta^{++}}^2 \Leftrightarrow {\delta^{0}}^2)$:

$$
\left(
\begin{array}{*3{C{1.5cm}}}
\la/4 & \frac{\la_1+\la_4}{2}\\ \cline{1-1}
\bord & \la_2+\la_3 \\ \cline{2-2}
\end{array}
\right).
$$

\underline{Copositivity  condition}: \\
$$ \la \geq 0, \hskip 1cm \la_2+\la_3 \geq 0, \hskip 1cm \la_1+\la_4 + \sqrt{\la\;(\la_2+\la_3)}  \geq 0. $$

\item
\begin{align}
^{2F}V_4(\phi^0\,,\,\delta^{++}) & = (\la_2+\la_3) {\delta^{++}}^4 + (\la_1) {\delta^{++}}^2\,{\phi^0}^2 + \frac{\la}{4} {\phi^+}^4. \\ \nonumber \\
^{2F}V_5(\phi^+\,,\,\delta^{0}) & = (\la_2+\la_3) {\delta^{0}}^4 + \la_1 {\phi^+}^2 {\delta^{0}}^2 + \frac{\la}{4} {\phi^+}^4.
\end{align}

In matrix form each can be represented in basis $({\phi^0}^2 \Leftrightarrow {\phi^+}^2, {\delta^{++}}^2 \Leftrightarrow {\delta^{0}}^2)$:

$$
\left(
\begin{array}{*4{C{1.5cm}}}
\la/4 & \frac{\la_1}{2}\\ \cline{1-1} 
\bord & \la_2+\la_3 \\ \cline{2-2}
\end{array}
\right).
$$

\underline{Copositivity  condition}: \\
$$ \la \geq 0, \hskip 1cm \la_2+\la_3 \geq 0, \hskip 1cm \la_1 + \sqrt{\la\, (\la_2+\la_3)}  \geq  0. $$

\item
\be
^{2F}V_6(\phi^+\,,\,\delta^{+})  = (\la_2+\frac{\la_3}{2}) {\delta^{+}}^4 + (\la_1+\frac{\la_4}{2}) {\delta^{+}}^2\,{\phi^+}^2 + \frac{\la}{4} {\phi^+}^4.
\ee

\be
^{2F}V_7(\phi^0\,,\,\delta^{+})  = (\la_2+\frac{\la_3}{2}) {\delta^{+}}^4 + (\la_1+\frac{\la_4}{2}) {\delta^{+}}^2\,{\phi^0}^2 + \frac{\la}{4} {\phi^0}^4.
\ee

In matrix form both of them can be represented in basis $({\phi^+}^2 \Leftrightarrow {\phi^0}^2, {\delta^{+}}^2)$:

$$
\left(
\begin{array}{*4{C{1.5cm}}}
\la/4 & \frac{\la_1+\frac{\la_4}{2}}{2}\\ \cline{1-1} 
\bord &\la_2+\frac{\la_3}{2} \\ \cline{2-2}
\end{array}
\right).
$$

\underline{Copositivity  condition}: \\
$$ \la \geq 0, \hskip 1cm \la_2+\frac{\la_3}{2}  \geq  0, \hskip 1cm \la_1+\frac{\la_4}{2}+ \sqrt{\la\, (\la_2+\frac{\la_3}{2})} \geq 0. $$

%
%


\item
\begin{align}
^{2F}V_8(\delta^+\,,\,\delta^{++}) & = (\la_2+\frac{\la_3}{2}) {\delta^{+}}^4 + 2 (\la_2+\la_3) {\delta^{+}}^2\,{\delta^{++}}^2 + (\la_2+\la_3) {\delta^{++}}^4.    \\ \nonumber \\
^{2F}V_{9}(\delta^0\,,\,\delta^{+}) & = (\la_2+\la_3) {\delta^{0}}^4 + 2 (\la_2+\la_3) {\delta^0}^2 {\delta^{+}}^2 + (\la_2+\frac{\la_3}{2}) {\delta^{+}}^4.
\end{align}

In matrix form each can be represented in basis $({\delta^+}^2 \Leftrightarrow {\delta^0}^2, {\delta^{++}}^2 \Leftrightarrow {\delta^{+}}^2)$:

$$
\left(
\begin{array}{*7{C{1.5cm}}}
\la_2+\la_3 & \la_2+\la_3\\ \cline{1-1}
\bord &\la_2+\frac{\la_3}{2} \\ \cline{2-2}
\end{array}
\right).
$$

\underline{Copositivity  condition}: \\
$$ \la_2+\la_3 \geq 0, \hskip 1cm \la_2+\frac{\la_3}{2}  \geq  0. $$


\item
\be
^{2F}V_{10}(\delta^0\,,\,\delta^{++})  = (\la_2+\la_3) {\delta^{0}}^4 + 2 \la_2 {\delta^0}^2 {\delta^{++}}^2 + (\la_2+\la_3) {\delta^{++}}^4.
\ee

In matrix form it can be represented  in basis $({\delta^{0}}^2, {\delta^{++}}^2)$:

$$
\left(
\begin{array}{*4{C{1.5cm}}}
\la_2+\la_3 & \la_2\\ \cline{1-1} 
\bord &\la_2+\la_3 \\ \cline{2-2}
\end{array}
\right).
$$

\underline{Copositivity  condition}: \\
$$ \la_2+\la_3  \geq  0,  \hskip 1cm \la_2 + \frac{\la_3}{2}  \geq 0. $$
 
\end{enumerate}


 \subsection{3-Field Directions and Stability Conditions}\label{app:vac_Type_II_3field}

\begin{enumerate}[(I)]

\item
\be
^{3F}V_1 (\delta^{0}\,,\,\delta^{+}\,,\,\delta^{++}) = \la_2\left ( {\delta^{0}}^2 +{\delta^{+}}^2 +{\delta^{++}}^2\right)^2 + \la_3 \left({\delta^{0}}^2 + \frac{{\delta^{+}}^2}{2}\right)^2+\la_3 \left(\frac{{\delta^{+}}^2}{2} + {\delta^{++}}^2\right)^2  
\ee

In matrix form both can be represented in basis $({\delta^0}^2, {\delta^{+}}^2, {\delta^{++}}^2)$:

$$
\left(
\begin{array}{*4{C{1.5cm}}}
\la_2+\la_3 & \la_{2}+\frac{\la_3}{2}& \la_2\\ \cline{1-1} 
\bord & \la_{2}+\frac{\la_3}{2} & \la_{2}+\frac{\la_3}{2} \\ \cline{2-2}
& \bord & \la_2+\la_3 \\ \cline{3-3}
\end{array}
\right).
$$

\underline{Copositivity  condition}: \\

$$ \la_2+\la_3  \geq  0,\hskip 0.8cm \la_2+\frac{\la_3}{2}  \geq  0, $$

\item
\begin{align} \label{3FV2_tII}
^{3F}V_2 (\phi^0\,,\,\delta^{+}\,,\,\delta^{0}) &= (\la_2+\la_3) {\delta^{0}}^4 +(\la_2+\frac{\la_3}{2}) {\delta^{+}}^4+ \frac{\la}{4}\;{\phi^0}^4 + 
2 (\la_2+\la_3) {\delta^0}^2 {\delta^{+}}^2\nonumber\\  & + (\la_1+\la_4) {\phi^0}^2 {\delta^{0}}^2 + (\la_1+\frac{\la_4}{2}) {\delta^{+}}^2\,{\phi^0}^2. \\ \nonumber \\  \label{3FV3_tII}
^{3F}V_3 (\phi^+\,,\,\delta^{+}\,,\,\delta^{++}) &= (\la_2+\frac{\la_3}{2}) {\delta^{+}}^4 +(\la_2+\la_3) {\delta^{++}}^4+  \frac{\la}{4}\;{\phi^+}^4 + 
2 (\la_2+\la_3) {\delta^+}^2 {\delta^{++}}^2 \nonumber\\ & + (\la_1+\la_4) {\phi^+}^2 {\delta^{++}}^2 + (\la_1+\frac{\la_4}{2}) {\delta^{+}}^2\,{\phi^0}^2.
\end{align}

In matrix form both can be represented in basis $({\phi^0}^2 \Leftrightarrow {\phi^+}^2, {\delta^+}^2, {\delta^{0}}^2 \Leftrightarrow {\delta^{++}}^2)$:

$$
\left(
\begin{array}{*4{C{1.5cm}}}
\frac{\la}{4}& \frac{\la_{1}+\la_4}{2} & \frac{\la_1+\frac{\la_4}{2}}{2}\\ \cline{1-1} 
\bord & \la_2+\la_3            & \la_2+\la_3 \\ \cline{2-2}
& \bord & \la_2+\frac{\la_3}{2} \\ \cline{3-3}
\end{array}
\right).
$$

\underline{Copositivity  condition}: \\

\begin{align} \nonumber
&\la  \geq  0, \hskip 0.8cm \la_2+\la_3  \geq  0,\hskip 0.8cm \la_2+\frac{\la_3}{2}  \geq  0, \\ \nonumber
&\kappa_1 = \la_2+\la_3  + \sqrt{(\la_2+\frac{\la_3}{2})  \left(\la_2+\frac{\la_3}{2}\right)}  \geq 0, \\ \nonumber
&\kappa_2 =  \frac{\la_1+\frac{\la_4}{2}}{2}+ \sqrt{\frac{\la}{4} (\la_2+\la_3)}  \geq 0 , \\ \nonumber
&\kappa_3 = \frac{\la_1+\la_4}{2} + \sqrt{\frac{\la}{4} (\la_2+\la_3)}  \geq 0, 
\end{align}

\beas
 \sqrt{\frac{\la}{4} \, (\la_2+\la_3) \, (\la_2+\frac{\la_3}{2})}&+& \frac{\la_{1}+\la_4}{2} \sqrt{\la_2+\frac{\la_3}{2}} 
 + \frac{\la_1+\frac{\la_4}{2}}{2}\sqrt{\la_2+\la_3} \\ &&  + (\la_2+\la_3)\sqrt{\frac{\la}{4}} + 
 \sqrt{2\,(\kappa_1)(\kappa_2)(\kappa_3)}  \geq  0.
\eeas

\item
\begin{align}
^{3F}V_4 (\phi^+\,,\,\delta^{+}\,,\,\delta^{0}) & =  (\la_2+\la_3) {\delta^{0}}^4 +(\la_2+\frac{\la_3}{2}) {\delta^{+}}^4+ \frac{\la}{4}\;{\phi^+}^4 + 
2 (\la_2+\la_3) {\delta^0}^2 {\delta^{+}}^2\nonumber \\  & + \la_1 {\phi^0}^2 {\delta^{0}}^2 + (\la_1+\frac{\la_4}{2}) {\delta^{+}}^2\,{\phi^+}^2. \\ \nonumber \\
^{3F}V_5 (\phi^0\,,\,\delta^{+}\,,\,\delta^{++}) & = (\la_2+\frac{\la_3}{2}) {\delta^{+}}^4 +(\la_2+\la_3) {\delta^{++}}^4+ \frac{\la}{4}\;{\phi^0}^4 + 
2 (\la_2+\la_3) {\delta^+}^2 {\delta^{++}}^2 \nonumber\\ & + \la_1 {\phi^0}^2 {\delta^{++}}^2 + (\la_1+\frac{\la_4}{2}) {\delta^{+}}^2\,{\phi^0}^2.
 \end{align}

In matrix form both can be represented in basis $({\phi^+}^2 \Leftrightarrow {\phi^0}^2, {\delta^+}^2,  {\delta^{0}}^2 \Leftrightarrow {\delta^{++}}^2)$:

$$
\left(
\begin{array}{*4{C{1.5cm}}}
\frac{\la}{4}       & \frac{\la_1}{2}    & \frac{\la_1+\frac{\la_4}{2}}{2} \\ \cline{1-1}
\bord    & \la_2+\la_3        & \la_2+\la_3 \\ \cline{2-2}
& \bord  & \la_2+\frac{\la_3}{2}\\ \cline{3-3} 
\end{array}
\right).
$$

\underline{Copositivity  condition}: \\

\begin{align} \nonumber
& \la  \geq  0, \hskip 0.8cm \la_2+\la_3  \geq  0,\hskip 0.8cm \la_2+\frac{\la_3}{2}  \geq  0,  \\ \nonumber
& \kappa_1 =  \frac{\la_1}{2}+ \sqrt{\frac{\la}{4} (\la_2+\la_3)}  \geq 0 ,  \\ \nonumber
& \kappa_2 = \frac{\la_1+\la_4}{2} + \sqrt{\frac{\la}{4} (\la_2+\la_3)}  \geq 0,  \\ \nonumber
& \kappa_3 = \la_2+\la_3  + \sqrt{(\la_2+\frac{\la_3}{2})  (\la_2+\la_3)}  \geq 0, 
\end{align}

\beas
\sqrt{\left(\frac{\la}{4}\right)\left(\la_2+\la_3\right)\left(\la_2+\frac{\la_3}{2}\right)}&+&\frac{\la_1+\frac{\la_4}{2}}{2}\sqrt{\la_2+\la_3}+\frac{\la_1}{2}\sqrt{\la_2+\frac{\la_3}{2}} \\ &
+& \la_2+\la_3\sqrt{\frac{\la}{4}} + \sqrt{2(\kappa_1)(\kappa_2)(\kappa_3)}  \geq  0.
\eeas
   
\item
\bea
^{3F}V_6 (\phi^0\,,\,\delta^{0}\,,\,\delta^{++}) &=& (\la_2+\la_3) {\delta^{0}}^4 +(\la_2+\la_3) {\delta^{++}}^4+ \frac{\la}{4}\;{\phi^0}^4 + 
2 \la_2 {\delta^0}^2 {\delta^{++}}^2\nonumber \\ &&+ \la_1 {\phi^0}^2 {\delta^{++}}^2 + (\la_1+\la_4) {\delta^{0}}^2\,{\phi^0}^2.
\eea

\bea
^{3F}V_7 (\phi^+\,,\,\delta^{0}\,,\,\delta^{++}) &=& (\la_2+\la_3) {\delta^{0}}^4 +(\la_2+\la_3) {\delta^{++}}^4+ \frac{\la}{4}\;{\phi^+}^4 + 
2 \la_2 {\delta^0}^2 {\delta^{++}}^2  \nonumber \\ &&+ \la_1 {\phi^+}^2 {\delta^{0}}^2 + (\la_1+\la_4) {\delta^{++}}^2\,{\phi^0}^2.
\eea

In matrix form both of them can be represented in basis $({\phi^0}^2 \Leftrightarrow {\phi^+}^2,  {\delta^0}^2,  {\delta^{++}}^2)$:

$$
\left(
\begin{array}{*4{C{1.5cm}}}
\frac{\la}{4} & \frac{\la_{1}+\la_4}{2} & \frac{\la_1}{2}\\ \cline{1-1}
\bord         & \la_2+\la_3   & \la_2 \\ \cline{2-2}
&\bord & \la_2+\la_3 \\ \cline{3-3}
\end{array}
\right).
$$

\underline{Copositivity  condition}: \\

\begin{align} \nonumber
& \la  \geq  0, \hskip 0.8cm \la_2+\la_3  \geq  0,\hskip 0.8cm \la_2+\frac{\la_3}{2}  \geq  0,   \\ \nonumber
& \kappa_1 = \frac{\la_1}{2}+ \sqrt{\frac{\la}{4} (\la_2+\la_3)}  \geq 0 ,   \\ \nonumber
& \kappa_2 = \frac{\la_1+\la_4}{2} + \sqrt{\frac{\la}{4} (\la_2+\la_3)}  \geq 0,  \\ \nonumber
& \kappa_3 = \la_2  + \sqrt{\frac{\la}{4} (\la_2+\la_3)}  \geq 0, 
\end{align}


\beas
 \sqrt{(\frac{\la}{4})(\la_2+\la_3)(\la_2+\la_3)} &+& \frac{\la_{1}+\la_4}{2}\sqrt{\la_2+\la_3}  + \frac{\la_1}{2} \sqrt{\la_2+\la_3 } 
 \\ &+& \la_2 \sqrt{\frac{\la}{4}} +  \sqrt{2(\kappa_1)(\kappa_2)(\kappa_3)} \geq 0.
\eeas


\item
\bea
^{3F}V_8 (\phi^0\,,\,\phi^{+}\,,\,\delta^{0}) &=& (\la_2+\la_3) {\delta^{0}}^4 +\frac{\la}{4}\;{\phi^0}^4+ \frac{\la}{4}\;{\phi^+}^4 + 
\frac{\la}{2} {\phi^+}^2 {\phi^{0}}^2  \nonumber\\ && + \la_1 {\phi^+}^2 {\delta^{0}}^2 + (\la_1+\la_4) {\delta^{0}}^2\,{\phi^0}^2.
\eea

\bea
^{3F}V_{9} (\phi^0\,,\,\phi^{+}\,,\,\delta^{++}) &=& (\la_2+\la_3) {\delta^{++}}^4 + \frac{\la}{4}\;{\phi^0}^4 +  \frac{\la}{4}\;{\phi^+}^4 + 
\frac{\la}{2} {\phi^+}^2 {\phi^{0}}^2  \nonumber\\ && + \la_1 {\phi^0}^2 {\delta^{++}}^2 + (\la_1+\la_4) {\delta^{++}}^2\,{\phi^+}^2.
\eea

In matrix form both can be represented in basis $({\phi^0}^2, {\phi^+}^2, {\delta^{0}}^2 \Leftrightarrow {\delta^{++}}^2)$:
$$
\left(
\begin{array}{*4{C{1.5cm}}}
\frac{\la}{4}       &\frac{\la}{4} & \frac{\la_1+\la_4}{2}\\ \cline{1-1} 
\bord &\frac{\la}{4}        & \frac{\la_1}{2} \\ \cline{2-2}
&\bord  & \la_2+\la_3 \\ \cline{3-3}
\end{array}
\right).
$$

\underline{Copositivity conditions}: \\

\begin{align} \nonumber
& \la  \geq  0, \hskip 0.8cm \la_2+\la_3  \geq  0, \\ \nonumber
& \kappa_1 = \frac{\la}{4}+ \sqrt{\frac{\la}{4}\cdot \frac{\la}{4}}  \geq 0 ,  \\ \nonumber
& \kappa_2 = \frac{\la_1+\la_4}{2} + \sqrt{\frac{\la}{4} (\la_2+\la_3)}  \geq 0,   \\ \nonumber
& \kappa_3 = \frac{\la_1}{2} + \sqrt{ (\la_2+\la_3) \frac{\la}{4}}  \geq 0,   
\end{align}

\beas
\sqrt{\left(\frac{\la}{4}\right)\left(\frac{\la}{4}\right)(\la_2+\la_3)}&+&\frac{\la_{1}+\la_4}{2}\sqrt{\frac{\la}{4}} 
+ \frac{\la_1}{2} \sqrt{\frac{\la}{4}} \\ &+& \frac{\la}{4} \sqrt{ \la_2+\la_3 } + \sqrt{2(\kappa_1)(\kappa_2)(\kappa_3)} \geq 0.
\eeas

\item
\bea \label{3FV10_tII}
^{3F}V_{10} (\phi^0\,,\,\phi^{+}\,,\,\delta^{+}) &=& (\la_2+\frac{\la_3}{2}) {\delta^{+}}^4 +\frac{\la}{4}\;{\phi^0}^4 + \frac{\la}{4}\;{\phi^+}^4 + 
 \frac{\la}{2} {\phi^+}^2 {\phi^{0}}^2 \nonumber \\ && + (\la_1+\frac{\la_4}{2}) {\delta^{+}}^2\,{\phi^+}^2 + (\la_1+\frac{\la_4}{2}) {\delta^{+}}^2\,{\phi^0}^2.
\eea

In matrix form it can be represented in basis $({\phi^0}^2, {\phi^+}^2, {\delta^{+}}^2)$:

$$
\left(
\begin{array}{*4{C{1.5cm}}}
\frac{\la}{4}    & \frac{\la}{4} & \frac{\la_1+\frac{\la_4}{2}}{2}\\ \cline{1-1}
\bord   & \frac{\la}{4}           & \frac{\la_1+\frac{\la_4}{2}}{2} \\ \cline{2-2}
        &\bord   & \la_2+\frac{\la_3}{2} \\ \cline{3-3}
\end{array}
\right).
$$

\underline{Copositivity conditions}: \\

\begin{align} \label{3FV10_tII:1} \nonumber
& \la  \geq  0, \hskip 0.8cm  \la_2+\frac{\la_3}{2}  \geq  0, \\ \nonumber
& \kappa_1 = \kappa_3 = \frac{\la_1+\frac{\la_4}{2}}{2} + \sqrt{\frac{\la}{4}\left( \la_2+\frac{\la_3}{2}\right) }  \geq  0, \\
& \kappa_2 =  \frac{\la}{4} + \sqrt{\frac{\la}{4}\cdot\frac{\la}{4}} =  \frac{\la}{2} \geq  0, 
\end{align}

\beas
\sqrt{\left(\frac{\la}{4}\right)\left(\frac{\la}{4}\right)\left(\la_2+\frac{\la_3}{2}\right)} &+& \frac{\la}{4} \sqrt{\la_2+\frac{\la_3}{2}}
+ \frac{\la_1+\frac{\la_4}{2}}{2}\sqrt{\frac{\la}{4}} 
\\ &+& \frac{\la_1+\frac{\la_4}{2}}{2}\sqrt{\frac{\la}{4}}+\sqrt{2(\kappa_1)(\kappa_2)(\kappa_3)} \geq 0.
\eeas

 
 \end{enumerate}

 \section{Conditions of {\sc cop}: LR Model with Triplet Scalars }\label{app:vac_LR_triplet_COP}

\subsection{2-Field Directions and Stability Conditions}\label{sec:vac_LR_triplet_2field}
\begin{enumerate}[(I)]

\item
\be
^{2F}V_1(\phi_1^0\,,\,\phi_1^+)  = \la_1 \big({\phi_1^0}^2 + {\phi_1^+}^2 \big)^2. \\
\ee

This can be represented as a symmetric matrix ($\Lambda$) of order two in basis $({\phi_1^0}^2, {\phi_1^+}^2)$:
 
$$
\left(
\begin{array}{*7{C{1.5cm}}}
\la_1 & \la_1\\  \cline{1-1}
\bord & \la_1 \\ \cline{2-2}
\end{array}
\right).
$$

\underline{Copositivity condition}: \\
$$ \la_1 \geq 0. $$
\item
\be
^{2F}V_2(\phi_1^0\,,\,\delta^{0}) = \la_5 \; {\delta^{0}}^4 + \la_1\;{\phi_1^0}^4. \\
\ee

\be
^{2F}V_3(\phi_1^+\,,\,\delta^{++}) = \la_5 \; {\delta^{++}}^4 + \la_1\;{\phi_1^+}^4. 
\ee

In matrix form both of them can be represented in basis $({\phi_1^0}^2 \Leftrightarrow {\phi_1^+}^2, {\delta^0}^2 \Leftrightarrow {\delta^{++}}^2)$:

$$
\left(
\begin{array}{*7{C{1.5cm}}}
\la_1 & 0\\  \cline{1-1}
\bord & \la_5 \\ \cline{2-2}
\end{array}
\right).
$$

\underline{Copositivity conditions}: \\
$$ \la_1 \geq 0 , \hskip 1cm \la_5 \geq 0. $$

\item
\begin{align} 
^{2F}V_4(\phi_1^0\,,\,\delta^{+}) &= (\la_5 + \la_6) {\delta^+}^4 + \la_1\;{\phi_1^0}^4 + \frac{1}{2}(\la_{12}+2\la_9)\;{\delta^{+}}^2 {\phi_1^0}^2.\\ \nonumber \\
^{2F}V_5(\phi_1^+\,,\,\delta^{+}) &= (\la_5 + \la_6) \; {\delta^+}^4 + \la_1\;{\phi_1^+}^4 + \frac{1}{2}(\la_{12}+2\la_9)\;{\delta^{+}}^2 {\phi_1^+}^2.
\end{align}

In matrix form both can be represented in basis $({\phi_1^0}^2 \Leftrightarrow {\phi_1^+}^2, {\delta^+}^2)$:

$$
\left(
\begin{array}{*7{C{2.2cm}}}
\la_1 & \frac{1}{4} (\la_{12}+2\,\la_9)\\ \cline{1-1} 
\bord & \la_5+\la_6 \\ \cline{2-2}
\end{array}
\right).
$$

\underline{Copositivity  condition}: \\
$$ \la_1 \geq 0 , \hskip 0.8cm \la_5+\la_6 \geq 0.$$

\item
\be
^{2F}V_6(\phi_1^0\,,\,\delta^{++}) = \la_5\;{\delta^{++}}^4 + \la_1\;{\phi_1^0}^4 + \la_{12}\;{\delta^{++}}^2 {\phi_1^0}^2.
\ee

\be
^{2F}V_7(\phi_1^+\,,\,\delta^{0})  = \la_5 \; {\delta^0}^4 + \la_1\;{\phi_1^+}^4 +\la_{12}\;{\delta^{0}}^2 {\phi_1^+}^2.
\ee

In matrix form each of them can be represented in basis $({\phi_1^0}^2 \Leftrightarrow {\phi_1^+}^2, {\delta^{++}}^2 \Leftrightarrow {\delta^{0}}^2)$:

$$
\left(
\begin{array}{*7{C{1.5cm}}}
\la_1 & \frac{\la_{12}}{2}\\ \cline{1-1} 
 \bord & \la_5 \\ \cline{2-2}
\end{array}
\right).
$$

\underline{Copositivity conditions}: \\
$$ \la_1 \geq 0 , \hskip 0.8cm \la_5 \geq 0.$$

\item
\be
 ^{2F}V_8(\delta^0\,,\,\delta^{+}) = \la_5 \big({\delta^0}^2 + {\delta^{+}}^2 \big)^2 + \la_6\; {\delta^{+}}^4.
 \ee

\be
 ^{2F}V_{9}(\delta^+\,,\,\delta^{++}) = \la_5 \big({\delta^+}^2 + {\delta^{++}}^2 \big)^2 + \la_6\; {\delta^{+}}^4 .
\ee

In matrix form both can be represented in basis $({\delta^+}^2, {\delta^0}^2 \Leftrightarrow {\delta^{++}}^2)$:

$$
\left(
\begin{array}{*7{C{1.5cm}}}
\la_5 & \la_5\\ \cline{1-1}
\bord & \la_5+\la_6 \\ \cline{2-2}
\end{array}
\right).
$$

\underline{Copositivity  condition}: \\
$$ \la_5 \geq 0 , \hskip 0.8cm \la_5+\la_6 \geq 0. $$

\item
\be
^{2F}V_{10}(\delta^0\,,\,\delta^{++}) = \la_5 \big({\delta^0}^2 + {\delta^{++}}^2 \big)^2 + 4 \la_6\; {\delta^{+}}^2{\delta^{0}}^2.
\ee

In matrix form it can be represented in basis $({\delta^0}^2, {\delta^{++}}^2)$:

$$
\left(
\begin{array}{*7{C{1.5cm}}}
\la_5 & \la_5+2\la_6\\ \cline{1-1}
\bord & \la_5 \\ \cline{2-2}
\end{array}
\right).
$$

\underline{Copositivity conditions}: \\
$$ \la_5 \geq 0 , \hskip 0.8cm \la_5+\la_6 \geq 0. $$


\end{enumerate}


\subsection{3-Field Directions and Stability Conditions}\label{sec:vac_LR_triplet_3field}

\begin{enumerate}[(I)]

\item
\begin{align} 
^{3F}V_1 (\phi_1^0\,,\,\phi_1^+\,,\,\delta^{0}) &= \la_1\; \big({\phi_1^0}^2+{\phi_1^+}^2 \big)^2 +\la_5\;{\delta^0}^2 +\la_{12}\;{\delta^0}^2{\phi_1^0}^2. \\ \nonumber \\
^{3F}V_2(\phi_1^0\,,\,\phi_1^{+}\,,\,\delta^{++}) &= \la_1\;\big({\phi_1^0}^2+{\phi_1^+}^2 \big)^2+\la_5\;{\delta^{++}}^4+\la_{12}\;{\phi_1^0}^2{\delta^{++}}^2.
\end{align}

In matrix form both of them can be represented in basis $({\phi_1^0}^2, {\phi_1^+}^2, {\delta^0}^2 \Leftrightarrow {\delta^{++}}^2)$:

$$
\left(
\begin{array}{*7{C{1.5cm}}}
\la_1 & \la_1 & \frac{\la_{12}}{2}\\ \cline{1-1}
\bord & \la_1 & 0 \\ \cline{2-2}
& \bord & \la_5 \\ \cline{3-3}
\end{array}
\right).
$$

\underline{Copositivity conditions}: \\
$$ \la_1 \geq 0, \hskip 0.8cm \la_5 \geq 0. $$

\item
\be
^{3F}V_3 (\phi_1^0\,,\,\phi_1^{+}\,,\,\delta^{+}) = \la_1\; \big({\phi_1^0}^2+{\phi_1^+}^2 \big)^2 + (\la_5+\la_6){\delta^+}^4 
                                                    +\frac{1}{2}(\la_{12}+2\la_9)\;\big({\phi_1^0}^2+{\phi_1^+}^2 \big) {\delta^+}^2.
\ee

In matrix form it can be represented in basis $({\phi_1^0}^2, {\phi_1^+}^2, {\delta^{+}}^2)$:

$$
\left(
\begin{array}{*7{C{2.2cm}}}
\la_1 & \la_1 &  \frac{1}{4} (\la_{12}+2\,\la_9)\\ \cline{1-1}
\bord & \la_1 & \frac{1}{4} (\la_{12}+2\,\la_9) \\ \cline{2-2}
& \bord & \la_5+\la_6 \\ \cline{3-3}
\end{array}
\right).
$$

\underline{Copositivity  condition}: \\
$$ \la_1 \geq 0, \hskip 0.8cm \la_5+\la_6 \geq 0. $$


\item
\begin{align}
^{3F}V_4 (\phi_1^0\,,\,\delta^{0}\,,\,\delta^{+})  &= \la_1\;{\phi_1^0}^4+\la_5\; \big({\delta^0}^2+{\delta^{+}}^2\big)^2 + \la_6 {\delta^+}^4 +
                                                        \frac{1}{2}(\la_{12}+2\la_9)\;{\phi_1^0}^2 {\delta^+}^2.  \\ \nonumber \\
^{3F}V_5 (\phi_1^+\,,\,\delta^{++}\,,\,\delta^{+}) &=  \la_1\;{\phi_1^+}^4+\la_5\; \big({\delta^+}^2+{\delta^{++}}^2\big)^2 + \la_6 {\delta^
{+}}^4 +
                                                         \frac{1}{2}(\la_{12}+2\la_9)\;{\phi_1^+}^2 {\delta^{+}}^2.  
\end{align}

In matrix form both can be represented in basis $({\phi_1^0}^2 \Leftrightarrow {\phi_1^+}^2, {\delta^0}^2 \Leftrightarrow {\delta^{++}}^2, {\delta^+}^2)$:

$$
\left(
\begin{array}{*7{C{2.2cm}}}
\la_1 & 0 & \frac{1}{4} (\la_{12}+2\,\la_9)\\  \cline{1-1}
\bord & \la_5 & \la_5 \\ \cline{2-2}
& \bord & \la_5+\la_6 \\ \cline{3-3}
\end{array}
\right).
$$

\underline{Copositivity  condition}: \\
$$ \la_1 \geq 0, \hskip 0.8cm \la_5 \geq 0,\hskip 0.8cm \la_5+\la_6 \geq 0. $$


\item
\bea  \label{eq:3FV6_1}
^{3F}V_6 (\phi_1^0\,,\,\delta^{0}\,,\,\delta^{++}) & =&  \la_5\; \big({\delta^0}^2+{\delta^{++}}^2\big)^2 + \la_1 \;{\phi_1^0}^4  + 4 \la_6
\; {\delta_0}^2{\delta^{++}}^2 \nonumber \\ 
& &  + \la_{12}\; {\delta^{++}}^2\, {\phi_1^0}^2 
             + 2\,\la_9\,\delta^0\,\delta^{++}\,{\phi_1^0}^2.
\eea

\bea  \label{eq:3FV6_2}
^{3F}V_7 (\phi_1^+\,,\,\delta^{0}\,,\,\delta^{++}) &=& \la_5\; \big({\delta^0}^2+{\delta^{++}}^2\big)^2 + \la_1 \;{\phi_1^+}^4 
                                                         + 4 \la_6\; {\delta_0}^2{\delta^{++}}^2 \nonumber \\
                                                     & &  + \; \la_{12}\; {\delta^0}^2\, {\phi_1^+}^2 + 2\,\la_9\,\delta^0\,\delta^{++}\,
{\phi_1^+}^2.
\eea

In matrix form both of them can be represented in basis $({\phi_1^0}^2 \Leftrightarrow {\phi_1^+}^2, {\delta^0}^2, {\delta^{++}}^2)$:

$$
\left(
\begin{array}{*7{C{3.5cm}}}
\la_1 & 0 & \frac{\la_{12}}{2}& \la_9 \\ \cline{1-1} 
\bord & \la_5 & (1-C)(\la_5+2\,\la_6) & 0  \\ \cline{2-2}
 & \bord & \la_5 & 0 \\ \cline{3-3}
 &  & \bord & 2C(\la_5+2\la_6) \\ \cline{4-4}
\end{array}
\right).
$$

\underline{Copositivity conditions}: \\
\bea  \label{eq:3FV6_c}
\la_1 \geq 0, \hskip 0.8cm \la_5 \geq 0,\hskip 0.8cm C(\la_5+2\,\la_6) \geq 0.
\eea

Here we encounter three possibilities in respect to the last condition for three ranges of the unphysical parameter $C$:  
\begin{enumerate}[(a)]
 \item \underline{ $ C  \geq  0 \;\mbox{such that} \; (1-C) \geq 0, $ i.e., $C \in [0,1]$:} 
\bea  \label{eq:3FV6_c1}
\la_5+2\,\la_6 \geq 0. 
\eea

 \item \underline{  $C  \geq  0 \;\mbox{such that} \; (1-C) < 0,$ i.e., $C \in (1,\infty]$:}
 \bea  \label{eq:3FV6_c2}
\la_5+2\,\la_6 \geq 0  \hskip 0.8cm \Band \hskip 0.8cm  \la_5^2 - (1-C)^2 (\la_5+2\,\la_6)^2 \geq 0.
\eea

\item \underline{ $C < 0\;\mbox{such that} \; (1-C) \geq 0,$ i.e., $C \in [-\infty,0)$:}
\bea  \label{eq:3FV6_c3}
\la_5+2\,\la_6 \leq 0  \hskip 0.8cm \Band \hskip 0.8cm  \la_5^2 - (1-C)^2 (\la_5+2\,\la_6)^2 \geq 0.
\eea
As $C$ is a free parameter we can rewrite this condition as:
\bea  \label{eq:3FV6_c3}
\la_5+2\,\la_6 \leq 0  \hskip 0.8cm \Band \hskip 0.8cm  \la_5^2 - C^2 (\la_5+2\,\la_6)^2 \geq 0,
\eea
with $C \in (1,\infty]$.
\end{enumerate}

 In principle, union of exhaustive scan over all possible $C$ values would provide us total allowed region 
 in the parameter space. That is indeed possible in much simpler way by finding the particular $C$ values which
 maximise the allowed region. Thus one can easily eliminate the unphysical parameter.
 Detailed discussion is in section~\ref{sec:vac_LR_triplet}. Similar method 
 would be implemented in many other cases as follows. 
 Thus the final  copositivity conditions in this case can be written as,
\bea  \label{eq:3FV6_final}
\la_1 \geq 0, \hskip 0.8cm \la_5 \geq 0,\hskip 0.8cm (\la_5+\la_6) \geq 0. 
\eea

\item
\bea
^{3F}V_8 (\phi_1^0\,,\,\delta^{+}\,,\,\delta^{++}) & = & \la_1\;{\phi_1^0}^4+\la_5\; \big({\delta^{++}}^2+{\delta^{+}}^2\big)^2 + \la_6\; {\delta^{+}}^4 \nonumber \\
          & & +   \frac{1}{2}\la_{12}\;{\phi_1^0}^2 (2{\delta^{++}}^2+{\delta^{+}}^2) + \la_9\,{\phi_1^0}^2\,{\delta^{+}}^2.
\eea

\bea
^{3F}V_9 (\phi_1^+\,,\,\delta^{0}\,,\,\delta^{+}) &=& \la_1\;{\phi_1^+}^4+\la_5\; \big({\delta^0}^2+{\delta^{+}}^2\big)^2 + \la_6\; {\delta^
{+}}^4 \nonumber \\
 & & +  \frac{1}{2}\la_{12}\;{\phi_1^+}^2 (2{\delta^0}^2+{\delta^{+}}^2)+ \la_9\,{\phi_1^+}^2\,{\delta^{+}}^2.
\eea

In matrix form both can be represented in basis $({\phi_1^0}^2 \Leftrightarrow {\phi_1^+}^2, {\delta^{++}}^2 \Leftrightarrow {\delta^{0}}^2, {\delta^+}^2)$:

$$
\left(
\begin{array}{*7{C{2.2cm}}}
\la_1  & \frac{1}{4} (\la_{12}+2\,\la_9) & \frac{\la_{12}}{2}\\ \cline{1-1} 
\bord  & \la_5+\la_6        & \la_5 \\ \cline{2-2}
 & \bord & \la_5 \\ \cline{3-3}
\end{array}
\right).
$$

\underline{Copositivity conditions}: \\

$$ \la_1 \geq 0, \hskip 0.8cm \la_5 \geq 0,\hskip 0.8cm \la_5+\la_6 \geq 0. $$

\item
\be
^{3F}V_{10} (\delta^0\,,\,\delta^{+}\,,\,\delta^{++}) =  \la_5\; \big({\delta^0}^2+{\delta^{+}}^2+{\delta^{++}}^2\big)^2 
                                                        +\la_6\; \big({\delta^{+}}^2+2\delta^0\delta^{++}\big)^2.
\ee

In matrix form it can be represented in basis $({\delta^0}^2, {\delta^+}^2, {\delta^{++}}^2)$:

$$
\left(
\begin{array}{*7{C{3cm}}}
\la_5 & \la_5 & (1-C)(\la_5+2\,\la_6) & 0 \\ \cline{1-1} 
\bord & \la_5+\la_6 & \la_5 & 2\la_6  \\ \cline{2-2}
 & \bord & \la_5 & 0 \\ \cline{3-3}
 &  & \bord & 2C(\la_5+2\la_6) \\ \cline{4-4}
\end{array}
\right).
$$

\underline{Copositivity  condition}: \\
$$ \la_1 \geq 0, \hskip 0.8cm \la_5 \geq 0,  \hskip 0.8cm \la_5+\la_6 \geq 0, \hskip 0.8cm C(\la_5+2\,\la_6) \geq 0.$$

Possible three cases are:  
\begin{enumerate}[(a)]
 \item \underline{$C > 0, \; (1-C) \geq 0, i.e., \; C\in [0:1]$} 
$$\la_5+2\,\la_6 \geq 0 \hskip 0.6cm \Band \hskip 0.6cm 2C\left(\la_5+\la_6\right)\left(\la_5+2\,\la_6\right) -4\,\la_6^2 \geq 0 $$

 \item \underline{$C > 0, \; (1-C) \leq 0, i.e., \; C\in [1:\infty]$}
$$\la_5+2\,\la_6 \geq 0 \hskip 0.6cm \Band \hskip 0.6cm \la_5^2 - (1-C)^2 (\la_5+2\,\la_6)^2 \geq 0. $$
$$2C\left(\la_5+\la_6\right)\left(\la_5+2\,\la_6\right) -4\,\la_6^2 \geq 0 $$

 \item \underline{$C > 0, \; (1-C) \leq 0, i.e., \; C\in [-\infty:0)$}
$$\la_5+2\,\la_6 \leq 0 \hskip 0.6cm \Band \hskip 0.6cm \la_5^2 - (1-C)^2 (\la_5+2\,\la_6)^2 \geq 0. $$
$$2C\left(\la_5+\la_6\right)\left(\la_5+2\,\la_6\right) -4\,\la_6^2 \geq 0 $$

\end{enumerate}

We have already discussed the similar situation in detail in the section~\ref{sec:vac_LR_triplet}.
Final conditions in this case can be calculated as:
$$ \la_1 \geq 0, \hskip 0.8cm \la_5 \geq 0,  \hskip 0.8cm \la_5+\la_6 \geq 0. $$

\end{enumerate}

\subsection{4-Field Directions and Stability Conditions}\label{sec:vac_LR_triplet_4field}

\begin{enumerate}[(I)]

\item
\bea
^{4F}V_{1} (\phi_1^0\,,\,\phi_1^+\,,\,\delta^{0}\,,\,\delta^{+}) & & = \la_5\; \big({\delta^0}^2+{\delta^{+}}^2\big)^2 +\la_6\;{\delta^{+}}^4
                                                         +\la_1\; \big({\phi_1^0}^2+{\phi_1^+}^2 \big)^2 \nonumber  \\
                                                       && + \frac{1}{2}\la_{12}\; \big(2{\delta^{0}}^2{\phi_1^+}^2 
                                                   + 2\sqrt{2}\phi_1^0\phi_1^+\delta^0\delta^{+}+   {\delta^+}^2\big({\phi_1^0}^2+{\phi_1^+}^2 \big)  \big) \nonumber \\ && +\la_9\,{\delta^+}^2\big({\phi_1^0}^2+{\phi_1^+}^2 \big) . 
\eea

In matrix form it can be represented in basis (${\phi_1^0}^2, {\phi_1^+}^2, {\delta^+}^2, {\delta^{++}}^2, \phi_1^0\,\phi_1^+, \delta^+\,\delta^{++}$):
$$
\left(
\begin{array}{*7{C{2.2cm}}}
\la_1     & (1-C)\la_1 & \frac{1}{4} (\la_{12}+2\,\la_9) & \frac{1}{4} (\la_{12}+2\,\la_9) & 0 & 0 \\ \cline{1-1}
\bord & \la_1     & 0 & \frac{\la_{12}}{2} & 0 & 0 \\ \cline{2-2}
 & \bord & \la_5+\la_6 & (1-K) \la_5 & 0 & 0 \\ \cline{3-3}
 &  & \bord & \la_5 & 0& 0 \\ \cline{4-4}
 &  &  & \bord & 2\,C \la_1 & \frac{\la_{12}}{\sqrt{2}} \\ \cline{5-5} 
 &  &  &  & \bord & 2\,K \la_5  \\ \cline{6-6}
\end{array}
\right).
$$

\underline{Copositivity  condition}: \\

$$ \la_1 \geq 0, \hskip 0.8cm \la_5 \geq 0, \hskip 0.8cm \la_5+\la_6 \geq 0, \hskip 0.8cm C \geq 0, \hskip 0.8cm K \geq 0. $$

Here the two possibilities are:
 
\begin{enumerate}[(I)]
 \item \underline{$(1-C) \leq 0, i.e., \; C \in (1,\infty]$}
$$ \la_1\;\la_1 - (1-C)^2\;\la_1^2 \geq 0  \hskip 0.8cm \Rightarrow \hskip 0.5cm  C \in [0,2]. $$

 \item \underline{$(1-K) \leq 0, i.e., \; K \in (1,\infty]$}
$$ \la_5\; (\la_5+\la_6) - (1-K)^2\;\la_5^2 \geq 0. $$ 
\end{enumerate}

The last condition leads to $\la_5+\la_6 \geq (1-K)^2 \la_5$, and this condition is maximally relaxed for $K=1$ as $\la_5 \geq 0$.  It is also possible to confirm numerically that $K=1$ allows most parameter space. So, $K=1$ and $C\in [0,2]$ are the possible choices.

Then final conditions are $$ \la_1 \geq 0, \hskip 0.8cm \la_5 \geq 0, \hskip 0.8cm \la_5+\la_6 \geq 0. $$

\item
\bea
^{4F}V_{2} (\phi_1^0\,,\,\phi_1^+\,,\,\delta^{0}\,,\,\delta^{++}) &=& \la_5\; \big({\delta^0}^2+{\delta^{++}}^2\big)^2 + 4\la_6\;{\delta^0}^2{\delta^{++}}^2
                                                                     +\la_1\; \big({\phi_1^0}^2+{\phi_1^+}^2 \big)^2 \nonumber  \\
                                                                      && + \la_{12}\left( \big( {\delta^{++}}^2{\phi_1^0}^2+{\delta^0}^2{\phi_1^+}^2 \big)
                                                                      +\delta^0 \, \delta^{++} \left({\phi_1^0}^2+{\phi_1^+}^2\right) \right)\nonumber \\ && +  2\la_9\,\delta^0 \, \delta^{++} \left({\phi_1^0}^2+{\phi_1^+}^2\right). 
\eea

In matrix form it can be represented in basis (${\phi_1^0}^2, {\phi_1^+}^2, {\delta^0}^2, {\delta^{++}}^2$):
$$
\left(
\begin{array}{*7{C{2.95cm}}}
\la_1     & \la_1 & 0 & \frac{\la_{12}}{2} & \la_9 \\ \cline{1-1}
\bord  & \la_1     & \frac{\la_{12}}{2} & 0 & \la_9 \\ \cline{2-2}
 & \bord & \la_5 & (1-C)(\la_5+2\,\la_6) & 0 \\ \cline{3-3}
 &  &\bord  & \la_5 & 0 \\ \cline{4-4}
 &  &  & \bord &  2C(\la_5+2\,\la_6) \\ \cline{5-5}
\end{array}
\right).
$$

\underline{Copositivity conditions}: \\
$$ \la_1 \geq 0, \hskip 0.8cm \la_5 \geq 0, \hskip 0.8cm  C(\la_5+2\,\la_6)  \geq  0.  $$

The possible three cases are:  
\begin{enumerate}[(a)]
 \item \underline{$C > 0, \; (1-C) \geq 0, i.e., \; C \in [0,1]$} 
$$ \la_5+2\,\la_6 \geq 0.  $$

 \item \underline{$C > 0, \; (1-C) \leq 0, i.e., \; C \in (1,\infty]$}
$$ \la_5+2\,\la_6 \geq 0  \hskip 0.8cm \Band \hskip 0.8cm  \la_5^2 - (1-C)^2 (\la_5+2\,\la_6)^2 \geq 0. $$

\item \underline{$C < 0, \; (1-C) \geq 0, i.e.,\; C \in [-\infty,0)$}
$$ \la_5+2\,\la_6 \leq 0  \hskip 0.8cm \Band \hskip 0.8cm  \la_5^2 - (1-C)^2 (\la_5+2\,\la_6)^2 \geq 0. $$
\end{enumerate}

We have already discussed the similar situation in detail in the section~\ref{sec:vac_LR_triplet}.
Final conditions in this case can be calculated as:
$$ \la_1 \geq 0, \hskip 0.8cm \la_5 \geq 0,  \hskip 0.8cm \la_5+\la_6 \geq 0$$

 \item
\bea
^{4F}V_{3} (\phi_1^0\,,\,\phi_1^+\,,\,\delta^{0}\,,\,\delta^{+}) &=&  \la_5\; \big({\delta^0}^2+{\delta^{+}}^2\big)^2 +\la_6\;{\delta^{+}}^4
                                                    +\la_1\; \big({\phi_1^0}^2+{\phi_1^+}^2 \big)^2 \nonumber \\
                                                  && + \frac{1}{2}\la_{12}\; \big(2{\delta^{0}}^2{\phi_1^+}^2 
                                           - 2\sqrt{2}\phi_1^0\phi_1^+\delta^0\delta^{+}+{\delta^+}^2\big({\phi_1^0}^2+{\phi_1^+}^2 \big)\big)\nonumber\\
                                                 && +\la_9\,{\delta^+}^2\big({\phi_1^0}^2+{\phi_1^+}^2 \big).
\eea

In matrix form it can be represented in basis (${\phi_1^0}^2, {\phi_1^+}^2, {\delta^+}^2, {\delta^{++}}^2, \phi_1^0\,\phi_1^+, \delta^+\,\delta^{++}$):
$$
\left(
\begin{array}{*7{C{2.2cm}}}
\la_1     &  (1-C) \la_1 & \frac{1}{4} (\la_{12}+2\,\la_9) & \frac{1}{4} (\la_{12}+2\,\la_9) & 0 & 0 \\  \cline{1-1}
  \bord & \la_1     & 0 & \frac{\la_{12}}{2} & 0 & 0 \\ \cline{2-2}
 & \bord & \la_5+\la_6 &  (1-K) \la_5 & 0 & 0 \\ \cline{3-3}
 && \bord & \la_5 & 0& 0 \\ \cline{4-4}
 &&& \bord & 2C \la_1 &- \frac{\la_{12}}{\sqrt{2}} \\ \cline{5-5} 
 &&&& \bord & 2K \la_5 \\ \cline{6-6}
\end{array}
\right).
$$

\underline{Copositivity  condition}: \\
$$ \la_1 \geq 0, \hskip 0.8cm \la_5 \geq 0, \hskip 0.8cm \la_5+\la_6 \geq 0, \hskip 0.8cm C \geq 0, \hskip 0.8cm K \geq 0, 
\hskip 0.8cm 4C \, K \, \la_1 \, \la_5 - \frac{\la_{12}^2}{2} \geq 0. $$

The two possible cases are:
 
\begin{enumerate}[(i)]
 \item \underline{$(1-C) \leq 0, i.e., C \in (1,\infty]$}
$$ \la_1\;\la_1 - (1-C)^2 \;\la_1^2 \geq 0  \hskip 0.8cm \Rightarrow \hskip 0.5cm  1 - (1-C)^2 \geq 0. $$

 \item \underline{$(1-K) \leq 0, i.e., K \in (1,\infty]$}
$$ \la_5\; (\la_5+\la_6) - (1-K)^2\;\la_5^2 \geq 0.  $$ 
\end{enumerate}

In a similar method discussed in detail in the section~\ref{sec:vac_LR_triplet},
we choose  $C = 2 $ and $K = 1$ for the last copositivity condition in this present case, e.g. 
$C \, K \, \la_1 \, \la_5 - \frac{\la_{12}^2}{8} \geq 0$. This choice of $C$ and $K$ are made 
keeping it in mind that these unphysical parameters can be set to values which allows most parameter space.

Here we can also argue the maximisation of the allowed parameter space, as suggested in section~\ref{sec:vac_LR_triplet}. As $C,K \geq 0$, we can rewrite the condition as 
$\la_1 \la_5 \geq |\la_{12}^2/(8CK)|$. Thus the largest parameter space can be accessed if we use the conditions $\la_1 \la_5 \geq 0$, which would be achieved for either $C$ or $K$ $\to$ $\infty$. But we have restriction on $C$ as
$0 \leq C \leq 2$. The other condition leads to $\la_5+\la_6 \geq (1-K)^2 \la_5$, and this condition is maximally relaxed for $K=1$ as $\la_5 \geq 0$. So as the product $\la_1 \la_5$ can be maximally relaxed for allowed maximum values of $C,K$ which are 2 and 1 respectively we find the following constraint on this product as 
$\la_1 \la_5 \geq \la_{12}^2/16$.

We finally arrived at the conditions as,
$$ \la_1 \geq 0, \hskip 0.8cm \la_5 \geq 0, \hskip 0.8cm \la_5+\la_6 \geq 0, \hskip 0.8cm 16\;\la_1\,\la_5 - \la_{12}^2 \geq 0. $$


 \item
\bea
 ^{4F}V_{4} (\phi_1^0\,,\,\delta^0\,,\,\delta^{+}\,,\,\delta^{++}) &=&  \la_5\; \big({\delta^0}^2+{\delta^{+}}^2+{\delta^{++}}^2\big)^2 
                                                                         +\la_6\; \big({\delta^{+}}^2+2\delta^0\delta^{++}\big)^2 + \la_1\;{\phi_1^0}^4 \nonumber  \\
                                                    & &  + \frac{1}{2}\la_{12}\;{\phi_1^0}^2 (2{\delta^0}^2+{\delta^{+}}^2)
                                          +\la_9{\phi_1^0}^2 \left({\delta^+}^2+2\,\delta^0\,\delta^{++}\right). 
\eea

\bea
 ^{4F}V_{5} (\phi_1^+\,,\,\delta^0\,,\,\delta^{+}\,,\,\delta^{++}) &=&  \la_5\; \big({\delta^0}^2+{\delta^{+}}^2+{\delta^{++}}^2\big)^2 
                                                                         +\la_6\; \big({\delta^{+}}^2+2\delta^0\delta^{++}\big)^2 + \la_1\;{\phi_1^+}^4 \nonumber \\
                                                                      & &   +\frac{1}{2}\la_{12}\;{\phi_1^+}^2 (2{\delta^0}^2+{\delta^{+}}^2)
                                                                           +\la_9{\phi_1^+}^2 \left({\delta^+}^2+2\,\delta^0\,\delta^{++}\right).
\eea

Both of them can be represented in basis (${\phi_1^0}^2 \Leftrightarrow {\phi_1^+}^2, {\delta^0}^2,{\delta^{+}}^2,{\delta^{++}}^2,\delta^0\,\delta^{++}$):

$$
\left(
\begin{array}{*6{C{2.95cm}}}
\la_1    & \frac{\la_{12}}{2} & \frac{\la_{12}}{4} & 0 & \la_9 \\ \cline{1-1}
 \bord & \la_5     & \la_5 & (1-C)(\la_5+2\,\la_6) & 0  \\ \cline{2-2}
& \bord & \la_5+\la_6 & \la_5  & 2\,\la_6 \\ \cline{3-3}
&& \bord & \la_5 & 0 \\ \cline{4-4}
&&& \bord & 2 C(\la_5+2\la_6)\\ \cline{5-5} 
\end{array}
\right).
$$

\underline{Copositivity conditions}: \\
$$ \la_1 \geq 0, \hskip 0.8cm \la_5 \geq 0, \hskip 0.8cm \la_5+\la_6 \geq 0, \hskip 0.8cm  C(\la_5+2\,\la_6) \geq 0.$$

The possible three cases are:  
\begin{enumerate}[(a)]
 \item \underline{$ C > 0, \; (1-C) \geq 0, i.e., \; C \in [0,1]$} 
$$ \la_5+2\,\la_6 \geq 0  \hskip 0.8cm \Band \hskip 0.8cm  2\,C (\la_5+\la_6) (\la_5+2\,\la_6) - 4\,\la_6^2 \geq 0. $$

 \item \underline{$C > 0, \; (1-C) \leq 0, i.e., \; C \in (1,\infty]$}
$$ \la_5+2\,\la_6 \geq 0  \hskip 0.8cm \Band \hskip 0.8cm  \la_5^2 - (1-C)^2 (\la_5+2\,\la_6)^2 \geq 0. $$
$$\hskip 0.8cm \Band \hskip 0.8cm  2\,C (\la_5+\la_6) (\la_5+2\,\la_6) - 4\,\la_6^2 \geq 0. $$

\item \underline{$C < 0, \; (1-C) \geq 0, i.e., \; C \in [-\infty,0)$}
$$ \la_5+2\,\la_6 \leq 0  \hskip 0.8cm \Band \hskip 0.8cm  \la_5^2 - (1-C)^2 (\la_5+2\,\la_6)^2 \geq 0, $$
$$\hskip 0.8cm \Band \hskip 0.8cm  2\,C (\la_5+\la_6) (\la_5+2\,\la_6) - 4\,\la_6^2 \geq 0. $$
\end{enumerate}

We have already discussed the similar situation in detail in the section~\ref{sec:vac_LR_triplet}.
Final conditions in this case can be calculated as:
$$ \la_1 \geq 0, \hskip 0.8cm \la_5 \geq 0,  \hskip 0.8cm \la_5+\la_6 \geq 0.$$

 \end{enumerate}



\section{Principal Sub-matrix approach} \label{app:Algo-PS}

\subsection*{Algorithm to examine copositivity of a order $n$ matrix}

Here we have demonstrated our principle of algorithm with a matrix of order $n$. 


Let us first define a matrix of order $n$ and some initialisation:

\begin{verbatim}
mats= {{a11,a12,...,a1n},{a21,a22,...,a2n},....,{an1,an2,...,ann}};
degree = Length[mats]; Print[degree];
mat[1, 1] = mats;
For[ii = 1, ii <= degree, ii++, {n[ii] = 1}];
matdummy[1, 1] = degree + 1;
mategsystm[1, 1] = Eigensystem[mat[1, 1]];
counter = 0;

\end{verbatim}

 Number of principal sub-matrices of matrix of order $n$ is $(2^n-1)$. It is very easy to identify the principal sub-matrices of order $n$ and $one$ of a matrix. It will have $n$-numbers of principal sub-matrices of order $one$ and they are the just diagonal elements of the original matrix. The matrix itself is the principal sub-matrix of order $n$.

\begin{verbatim}
For[ibig = 2, ibig <= degree,  ibig++, {
  For[ismall = 1, ismall <= Binomial[degree, ibig - 2], ismall++, {
    For[i[ibig] = 1, i[ibig] < matdummy[ibig - 1, ismall], i[ibig]++, {
      mat[ibig, n[ibig]++] = 
      Drop[mat[ibig - 1, ismall], {i[ibig]}, {i[ibig]}];
      matdummy[ibig, n[ibig] - 1] = i[ibig];
      mategsystm[ibig, n[ibig] - 1] = 
      Eigensystem[mat[ibig, n[ibig] - 1]];
      }]
    }]
  }]
\end{verbatim}

Now one needs to calculate all the eigenvalues. Next identify the negative eigenvalues.
Check whether the eigenvector associated with the negative eigenvalue is negative or not.
If the eigenvector is positive then the matrix of order $n$ is {\sl not} Copositive.

\begin{verbatim}
For[pp = 1, pp <= degree , pp++, {
  For[oo = 1, oo <= Binomial[degree, pp - 1], oo++, {
     For[ii = 1, ii <= degree + 1 - pp, ii++, {
        If[N[Extract[mategsystm[pp, oo], {1, ii}]] < 0, {
          vector[pp, oo] = N[Extract[mategsystm[pp, oo], {2, ii}]];
          If[MemberQ[vector[pp, oo], _?Positive] != 
                                    MemberQ[vector[pp, oo], _?Negative],
             Print["Error!!!!! ----->\tEigenvalue= ",
                          N[Extract[mategsystm[pp, oo], {1, ii}]], 
                          "\tEigenvector:\t", vector[pp, oo]]; counter++]
           }]
        }];
     }];
  }]

\end{verbatim}

Thus finally determine whether the matrix is copositive or not:

\begin{verbatim}
If[counter != 0, Print["\n The Matrix is NOT copositive."], 
                                 Print["\n The Matrix is copositive."]];
\end{verbatim}

\vskip 1cm

\subsection*{Numerical Example with an order four matrix} 
\begin{verbatim}
mats= {{1,-0.72,-0.59,0.6},  {-0.72,1,0.21,-0.46},  {-0.59,0.21,1,0.6},
                {0.6,-0.46,0.6,-1} } ;
\end{verbatim}

{\bf Principal sub-matrices of order $three$}

$$
\left(
\begin{array}{ccc}
1	& 0.21	& -0.46 \\
0.21 &	1 &	0.6\\
-0.46 &	0.6 &	-1
\end{array}
\right),
\left(
\begin{array}{ccc}
 1 & -0.72 & -0.59 \\
 -0.72 & 1 & 0.21 \\
 -0.59 & 0.21 & 1 \\
\end{array}
\right),$$
$$
\left(
\begin{array}{ccc}
 1 & -0.59 & 0.6 \\
 -0.59 & 1 & 0.6 \\
 0.6 & 0.6 & -1 \\
\end{array}
\right)
,
\left(
\begin{array}{ccc}
 1 & -0.72 & 0.6 \\
 -0.72 & 1 & -0.46 \\
 0.6 & -0.46 & -1 \\
\end{array}
\right).
$$
\vskip 0.6cm

\underline{Eigensystems associated with negative eigenvalues \{eigenvectors\}}
\\
\begin{verbatim}
-1.27588    {0.214546,-0.26745,0.939383}
-1.39819    {-0.300393,-0.300393,0.905278}
-1.19908    {-0.221234,0.129747,0.966551}
\end{verbatim}

\vskip 1cm

{\bf Principal sub-matrices of order $two$}

$$
\left(
\begin{array}{cc}
 1 & 0.6 \\
 0.6 & -1 \\
\end{array}
\right),
\left(
\begin{array}{cc}
 1 & -0.46 \\
 -0.46 & -1 \\
\end{array}
\right),
\left(
\begin{array}{cc}
 1 & 0.6 \\
 0.6 & -1 \\
\end{array}
\right),
$$

$$
\left(
\begin{array}{cc}
 1 & 0.21 \\
 0.21 & 1 \\
\end{array}
\right),
\left(
\begin{array}{cc}
 1 & -0.59 \\
 -0.59 & 1 \\
\end{array}
\right),
\left(
\begin{array}{cc}
 1 & -0.72 \\
 -0.72 & 1 \\
\end{array}
\right).$$
\vskip 0.6cm

\underline{Eigensystems associated with negative eigenvalues \{eigenvectors\}}
\\

\begin{verbatim}
-1.16619    {0.266934,-0.963715}
-1.10073    {-0.213904,-0.976855}
-1.16619    {0.266934,-0.963715}
\end{verbatim}   
\vskip 1cm

{\bf Principal sub-matrices of order $one$}
$$(1), (1), (1), (-1)$$
\vskip 0.6cm

\underline{Eigensystems associated with negative eigenvalues \{eigenvectors\}}
\\

\begin{verbatim}
-1    {1.}
\end{verbatim}   

   
 \begin{verbatim}
   The  matrix is NOT copositive.
 \end{verbatim}

 One of the diagonal element is negative and also in order two principal sub-matrix, one  principal sub-matrix has a negative eigenvalue associated with positive eigenvector. So the matrix is not copositive.


%

 \bibliography{vacuum_stability}
 
\end{document}